\documentclass[a4paper,12pt]{article}
\usepackage{amsfonts,amsthm,amsmath,amssymb,graphicx,hyperref,color}

\newcommand{\half}{{\textstyle {1 \over 2}}}

\def\vac{|0\rangle}
\def\bra#1{\left\langle #1\right|}
\def\ket#1{\left| #1\right\rangle}
\newcommand{\braket}[2]{\langle #1 | #2 \rangle}

\def\corr#1{\left\langle #1 \right\rangle}
\newcommand{\tr}{\operatorname{tr}}

\newcommand{\Sym}{\operatorname{Sym}}
\newcommand{\Com}{\operatorname{Com}}

\def\id{\textrm{id}}

\def\Dim{\textrm{Dim}}
\def\End{\textrm{End}}

\def\Sym{\textrm{Sym}}

\def\ra{\rangle}
\def\la{\langle}
\def\a{\alpha}
\def\b{\beta}

\def\m{\mu}

\def\s{\sigma}

\def\l{\lambda}
\def\L{\Lambda}

\def\d{\partial}

\def\C{\mathbb{C}}

\newcommand{\cN}{\mathcal N}
\newcommand{\cO}{\mathcal O}

\newcommand{\cR}{\mathcal R}

\newcommand{\cW}{\mathcal W}

\newcommand{\cZ}{\mathcal Z}

\newcommand{\be}{\begin{equation}}
\newcommand{\bea}{\begin{eqnarray}}

\newcommand{\ee}{\end{equation}}
\newcommand{\eea}{\end{eqnarray}}

\newcommand{\nn}{\nonumber}

\def\bI{\mathbb{I}}

\begin{document}

\rightline{QMUL-PH-08-12}

\vspace{2truecm}

\vspace{15pt}

%%%%%%%%%%%%%%%%%

\centerline{\LARGE \bf Diagonal free field matrix  correlators, } 
\centerline{ \LARGE \bf   global symmetries and giant gravitons }
 \vspace{1truecm}
\thispagestyle{empty} \centerline{
    {\large \bf T.W. Brown}\footnote{E-mail address:
                                  {\tt t.w.brown@qmul.ac.uk}},
    {\large \bf P.J. Heslop}\footnote{E-mail address:
                                  {\tt p.j.heslop@qmul.ac.uk}}
    {\bf and}
    {\large \bf S. Ramgoolam}\footnote{E-mail address:
                                  {\tt s.ramgoolam@qmul.ac.uk}}
                                                       }

\vspace{.4cm}
\centerline{{\it  Centre for Research in String Theory, Department of Physics}}
\centerline{{ \it Queen Mary, University of London}}
 \centerline{{\it Mile End Road, London E1 4NS, UK}}

\vspace{1.5truecm}

%%%%%%%%%%%%%%%%%
\thispagestyle{empty}

\centerline{\bf ABSTRACT}

\vspace{.5truecm}

\noindent
We obtain a basis of diagonal free field multi-matrix 2-point
correlators in a theory with global symmetry group $G$.  The operators
fall into irreducible representations of $G$.  This applies for gauge
group $U(N)$ at finite $N$.  For composites made of $n$ fundamental
fields, this is expressed in terms of Clebsch-Gordan coefficients for
the decomposition of the $n$-fold tensor products of the fundamental
field representation in terms of $ G \times S_n $ representations. We
use this general construction in the case of the $SL(2)$ sector of
$\cN=4$ SYM. In this case, by using oscillator constructions, we
reduce the computation of the relevant Clebsch-Gordans coupling
infinite dimensional discrete series irreps of $SL(2)$ to a problem in
symmetric groups.  Applying these constructions we write down gauge
invariant operators with a Fock space structure similar to that
arising in a large angular momentum limit of worldvolume excitations
of giant gravitons.  The Fock space structure emerges from Clebsch
multiplicities of tensor products of symmetric group representations.
We also give the action of the 1-loop dilatation operator of $\cN=4$ SYM 
on  this basis of multi-matrix operators.

\vspace{.5cm}

\newpage

\tableofcontents

\setcounter{footnote}{0}

%---------------------------------------------------------------------------

\section{Introduction}

AdS/CFT \cite{malda,gkp,wit} 
allows a translation of many  hard questions of 
quantum gravity to questions in the dual CFT. 
Intriguing lessons on quantum gravity have taken the form 
of the stringy exclusion principle \cite{malstrom}. Some aspects 
of this principle find a geometrical expression in the properties 
of giant gravitons \cite{mst,giantgravitons}.  The systematic study 
of the correlation functions of gauge invariant multi-trace local  operators 
in the half-BPS sector has been fruitful in identifying CFT duals 
\cite{bbns,cjr,bertoy}  of giant gravitons, 
strings attached to them \cite{bbfh,rob,bcv}    
as well as bulk geometries resulting from their back-reaction 
on space-time \cite{llm}. 
An important step in these maps to spacetime is the identification 
of appropriate diagonal bases in the space of multi-trace 
operators. For holomorphic operators constructed from a single 
complex matrix, relevant to the half-BPS sector, 
this was solved in \cite{cjr}. Recent 
generalizations including non-holomorphic operators and 
multi-matrices have been achieved \cite{kimram,bhr,bm}.

In \cite{bhr} we solved the diagonalisation in the 
sector of holomorphic operators constructed from $M$  
complex matrices transforming in the fundamental 
of $U(M)$ (the case of $M=3$ being of interest in 
 $\cN = 4 $ SYM). We gave  covariant operators with correlators 
of the simple form
\begin{equation}
  \label{cov}  
\corr{ (  \hat  \cO^{\Lambda, M_\L ,i } )^{I}_J    
(\hat \cO^\dagger{}^{\Lambda' ,  M'_{\L'}, i'} )^K_L } 
= \delta^{{ \L} {\L}' }  \delta^{ M_\L M'_{\L'} } 
\sum_{\s \in S_n}    ~ D^{{\L} }_{i i' } ( \s) ~~  
    ( \s )_{J}^{K}   ( \s^{-1}   )^{I}_{L}
\end{equation}
We will call this the {\it canonical covariant form}. $\L$ is the
$U(M)$ irreducible representation and $M_\L$ labels the state within
this irrep.  $D^{\L}_{ii'}(\s)$ is the orthogonal matrix
representation of the $S_n$ representation $\L$.  There is no
spacetime dependence because we are considering a 4d analogue of the
Zamolodchikov metric used in 2d CFT. We will often use the 
expression `two-point function' interchangeably with this metric. 

These covariant operators lead to gauge invariant operators with
diagonal correlation functions
\begin{equation}
 \corr{ \cO^{\L, M_\L  ,R, \tau }
 \cO^\dagger{}^{\L'   , M'_{\L'},R', \tau' } } \propto 
\delta^{{\L} {\L}' } \delta^{ M_\L M'_{\L'} }  \delta^{R R' } \delta^{ \tau \tau' }
    \label{orthoresult} 
\end{equation}
$R$ is the $U(N)$ representation which organises the multi-trace
structure of the operator.

In this paper we will explain how to get the canonical covariant
form~(\ref{cov}) for any global symmetry group $G$. This will use
Schur-Weyl duality which we review below.  The diagonal operators for
the gauge invariant operators (\ref{orthoresult}) follows
automatically (see Section \ref{sec:gaugeN}).

Applied to $\cN$=4 SYM this means we solve the problem of writing down
a basis for the space of all gauge invariant operators in the theory
(and in any subsector of the theory).  Operators are labelled by
lowest weight representations of the global symmetry group and
diagonalise the free two-point correlation functions. We write these
operators in terms of appropriate Clebsch-Gordon coefficients and it
should be noted that finding the CG coefficients themselves is still a
difficult problem in general, but we show how to do this in the
particular example of the $SL(2)$ sector.

\subsection{Schur-Weyl duality and $G \times S_n$ Clebsch decompositions }

Let $V_F$ be the fundamental representation of $U(M)$ (or $GL(M)$).  
Classical Schur-Weyl duality gives the decomposition of 
$V_F^{ \otimes n } $ in terms of irreducible representations of $U(M)$.
 It relies on the fact 
that in the  algebra of linear  operators acting $V_F^{ \otimes n } $, 
i.e. in $\End ( V_F^{ \otimes n } ) $, the maximal subalgebra which commutes with 
$U(M)$ is exactly the group algebra of $S_n$. It gives the decomposition 
of $ V_F^{\otimes n } $ under $ U(M) \otimes S_n $ :   
\begin{align}
  \label{swd}
  V_F^{ \otimes n } = \bigoplus_{ \L \vdash n }~~  V_{ \L }^{ U(M) } \otimes
  V_{\L}^{ S_n }
\end{align}
Here $ \L \vdash n $ denotes the fact that $\L$ runs over partitions 
of $n$, which correspond to Young diagrams with row lengths  
$( r_1, r_2 \cdots )$ 
with $ r_i \ge r_{i+1} $. According to (\ref{swd}),  $ V_F^{ \otimes n }$ 
 has a complete basis of states of the form 
$ | \L , M_{ \L } , m_{ \L } \ra $, where $M_{\L}$ label states 
in the irrep. of $U(M)$ corresponding to the Young diagram $\L$ 
and $m_{\L}$ label states in the irrep. of $S_n$ corresponding to the 
same Young diagram.  
For $n > M $, the above has to be qualified with the 
constraint that $ c_1 ( \L ) \le M $, i.e. the maximal   length
of the first column of the Young diagram is $M$.  

Schur-Weyl duality is a special case of the double centraliser theorem
(or double commutant theorem) 
which gives a decomposition generalising  (\ref{swd}) for any algebra
acting on a vector space $\mathcal{ W }  $ (for a brief statement of 
the key relevant facts see Section 1 of \cite{halverson} or textbooks such 
as \cite{goodwall}).
The decomposition is given in terms of 
the algebra of interest $A$ and the commutant (centraliser)
of its action in the vector space, often denoted as $End_A ( \cW )$.
Let $G$ be the global symmetry group of 
a field theory. Let $V_F$ be a representation of 
$G$ formed by a set of fields in the theory. 
We will consider  $ \cW = V_F^{\otimes n } $ and the algebra 
of interest is the universal enveloping algebra of the Lie algebra of
 $G$.  
Often $G$ will be a subgroup of the full symmetry group 
of a gauge theory. 
In the case of $\cN=4$ SYM, if we are interested in
the six hermitian scalars (without space-time derivatives), 
$V_F$ is the fundamental of $SO(6)$. If we  write the six hermitian 
scalars as three complex ones, and only consider the 
holomorphic combinations, then the symmetry group of interest is $U(3)$
and $V_F$ is the fundamental $3$-dimensional representation. This is
the  case we studied in~\cite{bhr} and is covered by classical 
Schur-Weyl duality.
In this paper one of the main examples will involve $ G = SL(2)$ 
and $V_F$ will be an infinite dimensional discrete series 
representation corresponding to a single scalar field $ X $ and its
derivatives with respect to one light-cone direction, $\d X , \d^2 X ,
 \cdots $.   More generally we have fields
  $W_{m}\in V_F$ where $m$ may run over an
infinite dimensional vector space. In the $SL(2)$ example we would
have $W_m=\d^m \Phi$. We will then consider the $n$-field composites 
\bea
  W_{m_1} \otimes W_{m_2} \otimes \cdots \otimes W_{m_n}
\eea
which transform in the representation $V_F^{\otimes n}$ of $G$. 
This space also has an action of  the symmetry group $S_n$ which
permutes  the $n$ fields. This $S_n$ action commutes with 
$G$, but does not provide the full commutant of $G$. 
Equivalently there is a non-trivial commutant algebra of $ G \times S_n$, 
denoted by  $ \Com ( G \times S_n ) $.  By the double-commutant theorem, 
this will organise the multiplicities of $ G \times S_n $ representations.  
We have 
\begin{equation}\label{gsn}
   (V_F^G)^{\otimes n}  = \bigoplus_{\L, \L_1} V_{\L}^G \otimes
   V_{\L_1}^{S_n} \otimes V_{\L, \L_1}^{ \Com ( G \times S_n ) }
\end{equation}
where $\L$ is a representation of the global group $G$ and $\L_1$ is
a representation of $S_n$.  $V_{\L, \L_1}$ gives the multiplicity
with which $\L$ of $G$ and $\L_1$ of $S_n$ appear together.
It is a representation of $ \Com ( G \times S_n ) $. The explicit form 
of   $ \Com ( G \times S_n ) $ will not be needed in this paper,  
with  $V_{\L, \L_1}$ appearing simply as the carrier space of a 
multiplicity label. In  diagonalising the free-field two point functions,
the decomposition of  $V_F^{ \otimes n } $
 in terms of the group $ G \times S_n$ will be crucial.

The study of the multiplicity of 
the decomposition of $n$-fold tensor powers in terms of 
$ G \times S_n$ is called the plethysm problem. 
This  is solved for $SU(2)$ in \cite{kingpleth}
and  is, for more general $G$, the subject of a large mathematical 
literature.  For developments on the use of the combinatorics
related to plethysms in the context of chiral rings of 
a large class of $\cN =1$ SYM theories see \cite{hehan}.  
For our purposes, we will be interested, not only 
in the multiplicities of plethysms but 
also the corresponding Clebsch-Gordan coefficients. Using these
we will obtain explicit gauge invariant operators which diagonalize the 
Zamolodchikov metric.

\subsection{Outline of paper and main results}
In Section 2 we will find the canonical covariant form for the $SL(2)$
sector. In particular we will have in mind derivatives acting on one
complex scalar $ X $.  In Section 3, we will give explicit formulae
for the $SL(2) \times S_n$ multiplicities involved. We will be heavily
using the oscillator construction of $SL(2)$.  In Section 4, we show,
for any global symmetry $G$ and for $n$-field composites with the
basic fields transforming in any representation $V$ of $G$, that the
covariant canonical form follows once we construct $n$-field
composites using the Clebsch-Gordan coefficients for the $G \times
S_n$ decomposition of $ V^{ \otimes n } $. In Section 5 we give
various examples of this general case explaining how the
diagonalisations in the $ U(M)$ and $SL(2)$ sectors follow this
general pattern, and how the canonical covariant corresponding to
representations of the higher spin group $HS(1,1)$ are also included.  We describe some
useful facts concerning $ SO(6) 
\times S_n$ multiplicities relevant to the $SO(6)$ sector of six
hermitian scalars and finally we show how to apply the construction of
Section 4 to the case where where $G$ is of the product form $G_1
\times G_2$.  In Section 6 we review how to get the diagonal
gauge-invariant operators from the canonical covariant form and show
the compatibility of the counting with matrix model methods of
\cite{dutgop}.  Section 7 shows how our results on the $SL(2)$ sector
can be used to provide the gauge theory operators dual to the
worldvolume excitations of giants considered in \cite{djm}.  The
one-loop mixing of the gauge-invariant operators constructed in
Section 6 for $\cN= 4$ is analysed in Section 8.

\section{$SL(2)$  sector: covariant operators}\label{sec:SL2}

We consider the $SL(2)$ sector which we can view as a reduction of
$\cN=4$ SYM  to a sector with a single light-cone derivative of the complex scalar
$X$. We choose $\d \equiv (\d_{0}+\d_3)/2$.

We find the basic two-point function
\begin{equation}
  \corr{ \d^{k_1}X^{\dagger}{}^i_j(x)\; \d^{k_2} X^k_l(0)} = \frac{(-1)^{k_1} (k_1 +k_2+1)!   }{x^{2+k_1+k_2}}\;\delta^i_l\, \delta^k_j\label{eq:basic}
\end{equation}

If we consider $\cN=4$ SYM on $\cR^4$, we take our two operators to 
zero and infinity (corresponding to opposite poles of the 
conformally equivalent $S^4$) we have 
\begin{equation}\label{zammet}
  \corr{ \d^{k_1}X^{\dagger}{}'{}^i_j(x' = 0)\; \d^{k_2} X^k_l(x=0)}
  =\delta^{k_1k_2}  (k_1!)^2  \;\delta^i_l\, \delta^k_j
\end{equation}
where $x'=x/x^2$ is the coordinate patch around the north pole and $x$
around the south.
This technique is well known from the studies of conformal field
theories in two dimensions and the above  is
known as the Zamolodchikov metric (see \cite{ginsparg,polchinski} 
for a general account and \cite{copto} for applications to 
$ \cN = 4 $ SYM). 
Note that  this metric on operators is defined using space-time 
dependent two-point functions but is itself 
independent of spacetime. Knowing the metric for arbitrary derivatives 
allows a reconstruction of the spacetime dependence.

\subsection{Oscillator construction}\label{sec:osccon}

The oscillator representation allows an elegant method of
constructing primary fields in the $SL(2)$ sector~\cite{beisert,bianchi}.
By using this representation to find 
the Clebsch-Gordan coefficients associated with the 
$SL(2) \times S_n$, we will  solve the problem of finding 
the canonical covariant 2-point functions. 
It will turn out that in addition to the groups $SL(2)$ 
and $S_n$ another symmetric group 
will play an interesting role. It is $S_k$ where 
$k$ is the number of derivatives required to construct the 
lowest  weight state. 

The  $ SO(4,2) $ conformal algebra is given by
\begin{align} [M_{ab}, P_c ] = {}&   \eta_{bc} P_a- \eta_{ac} P_b
  \, , \qquad [M_{ab}, K_c ] =  \eta_{bc} K_a -  \eta_{ac} K_b \, ,
  \\ [M_{ab}, M_{cd}] = {}&  \eta_{bc} M_{ad} - \eta_{ac} M_{bd} +
  \eta_{ad} M_{bc}- \eta_{bd} M_{ac} \, , \\ [D,P_a] = {}&  P_a \,
  , \quad [D,K_a] = -  K_a \, , \qquad [K_a , P_b] = 
  2\eta_{ab} D- 2 M_{ab} \, 
\end{align}
The $SL(2)$ sector in terms of the conformal generators  
can be chosen as 
\begin{align}
  L_+=\half(P_0+P_3)\qquad L_-=\half(K_0-K_3)\qquad L_0=\half(D-M_{03})
\end{align}
giving
\begin{align}
  [ L_-,L_+ ]=2\,L_0\;, \qquad [L_0,L_{\pm} ]=\pm L_{\pm}
\end{align}

This algebra may be represented using  oscillators as
\begin{align}
  L_+=a^\dagger + a^\dagger a^\dagger a\;, \qquad L_0= \half+a^\dagger
  a\;, \qquad L_-=a\
\end{align}
where $[a,a^\dagger]=1$.
The lowest weight state of the representation $V_F$ is denoted $\vac$
and is annihilated by all the lowering oscillators $L_-=a$. It can
straightforwardly be checked that the 
raising operators $L_+$ then act on the lowest weight state as
\begin{align}
  (L_+)^k\,\vac=k! \,(a^\dagger)^k \,\vac 
\qquad \leftrightarrow \qquad  \d^k\, X\  | 0 \ra \label{remebmer}
\end{align}
By the operator-state correspondence, the  operator on the RHS above  
acts on the CFT vacuum at the origin in  radial quantization 
to give a state. Hence we have a map from oscillator states 
used in the representation theory of $SL(2)$ to states in radial 
quantization. Dual states in the oscillator Hilbert space 
map to states at the dual vacuum (at infinity) in radial quantization. 
\bea 
 \la 0 | L_-^k    = \la 0 | a^k   \qquad \leftrightarrow \qquad  \la 0 | ~ \d^k\, X\ 
\eea 
Note that the usual oscillator inner product correctly maps 
to the inner product given by the Zamolodchikov metric (\ref{zammet}).
 The normalization in   (\ref{zammet}) is easily calculated using the 
$SL(2) $ algebra once we use the fact that $L_- $ is the hermitian conjugate 
of $L_+ $ in radial quantization. 

In a similar way we can represent the tensor product $V_F^{\otimes n}$
by considering $n$ independent oscillators $a_i$. In this  space
the action of the diagonal  $SL(2)$  is obtained by summing over $n$:
\begin{align}
  {\bf L}_+=\sum_i (a_i^\dagger + a_i^\dagger a_i^\dagger a_i)\;, \qquad 
{\bf L}_0=
  \half n +\sum_i  a_i^\dagger
  a_i\;, \qquad {\bf L}_-=\sum_i a_i\label{bfL}
\end{align}
The relation between the oscillator states and the field states is:
\begin{align}\label{relox}
 \prod_{l=1}^n \, (a^\dagger_l)^{k_l}\, \vac \qquad \leftrightarrow \qquad
 {1\over k_1! k_2!\dots k_n!} \d^{k_1} X \,\otimes \,\d^{k_2} X \,\otimes \dots \otimes \,\d^{k_n} X  ~~ | 0 \ra 
\end{align}
The lowest weights are  annihilated by $L_-=\sum_i a_i$. All the lowest 
weight states at level $ L_0 = n+k $ are generated by 
$k$-oscillator states obtained as products of 
$ (a^\dagger_{i}-a^\dagger_{j}) $ acting on the vacuum. 
The simplest example is at $n=2$
where the lowest weight states are all of the form
\begin{align}
  \cO_k = (a^\dagger_1-a^\dagger_2)^k  \vac
\end{align}
Expanding out the oscillators and using~(\ref{relox}) we find the
corresponding operators in field space
\begin{align}
  \cO_k  \sim \sum_{j=0}^k {k \choose j}^2 (-1)^{k-j}\; \d^j X \otimes \d^{k-j} X
\end{align}
These are conformal higher spin currents, first constructed in
\cite{Konstein:2000bi}.

\subsection{Action of $S_n$ on $A_a$}

The lowest weight states  are
clearly generated by the $n-1$ differences of oscillators
 $a_i^{\dagger} - a_j^{\dagger} $.  The following basis 
will be very useful
\begin{align}
    A_a^{\dagger} = \sum_{ i =1 }^{n} J_a{}^i a_i^{\dagger}
\end{align}
where $J_a{}^i$ takes us from the natural representation of $S_n$ on
$n$ objects (labelled by the index $i$) to the $n-1$ dimensional
$H=[n-1,1]$ representation for which we will choose the orthonormal
basis (labelled by the index $a$)\footnote{The natural representation
  of $S_n$ on $n$ objects is a reducible representation and decomposes
  into the $n-1$ dimensional rep $H=[n-1,1]$ and the 1-dimensional
  symmetric rep $[n]$. We consider the map from $a_i$ onto the
  symmetric representation in Section~\ref{sec:descendants} where we
  see that they correspond to the application of the $SL(2)$
  descendant operator.\label{foot:natural}}.  The matrix $J$ will thus
have the following properties
\begin{align}
  D^H_{ab}(\sigma)J_b{}^i&=J_a{}^{\sigma(i)}\\
  J_a{}^i (J^\dagger)_b{}^i&=\delta_{ab}\ 
\end{align}
with $D^H_{ab}(\sigma)$ the orthonormal representation.
Explicitly we find
\begin{equation}\label{defA}
  A_a^\dagger = \frac{1}{\sqrt{i(i+1)}}\left(a_1^\dagger + \dots a_a^\dagger
    - ia_{a+1}^\dagger\right)\ 
\end{equation}
The details of the $S_n$ action on $A_a^\dagger$, and its relation to
the orthogonal representing matrix of the hook representation
$D^H_{ab}(\sigma)$, are given in Appendix Section
\ref{sec:matrixconstruct}.

\subsection{Canonical covariant form using oscillators} 

The simplest operator in the space $V_F^{\otimes n}$ is given by $n$
scalar fields without derivatives 
\bea 
 (\cO_n){}^{i_1 i_2 \cdots i_n }_{ j_1 j_2 \cdots j_n } 
 = X^{i_1}_{j_1} \otimes  X^{i_2}_{j_2} \otimes \cdots  \otimes X^{i_n}_{j_n}
\eea 
 The scalar fields lie in the adjoint representation of the
gauge group $U(N)$. Using multi-indices 
 $I = (i_1 , i_2 \cdots i_n) $  and $J = (j_1 , j_2 \cdots j_n ) $ 
we can write 
\bea 
(\cO_n){}^{I}_{J} &=& ( X \otimes X \otimes \cdots X )^I_J \equiv { \bf X }^I_J 
\eea  
There is a map which takes the states in the oscillator 
construction of the $V_F^{\otimes n  } $ representation of 
$SL(2)$ to operators or states (by  the operator-state correspondence)
in the CFT. The vacuum of the oscillator construction maps to 
$  { \bf X }^I_J  |0\ra $. Denoting the map from 
oscillator Hilbert space to the CFT states as $\rho^I_J  $ we may
write  
\bea 
  \rho^I_J (   |0 \ra )  \leftrightarrow  { \bf X }^I_J  |0\ra
     \equiv  |  { \bf X }^I_J \ra\ 
\eea
The vacuum on the right can be viewed as the vacuum in the CFT associated 
with  the origin in radial quantization.
The dual oscillator vacuum maps to the dual vacuum at infinity 
in radial quantization. 
\bea 
 \rho^I_J (   \la 0 | )  \leftrightarrow \la 0| { \bf X^{\dagger}  }^I_J 
 \equiv  \la { \bf X }^I_J |
\eea 
 The correlation function of these operators is
\begin{align}
 \braket{{\bf X}^I_J}{{\bf X}^K_L}=\sum_{\sigma\in S_n}
  \sigma^I_L (\sigma^{-1})^K_J\ 
\end{align}
Now the operator corresponding to $a_i^\dagger |0 \ra $ has $n-1$
scalar fields $X$ with one derivative of a scalar field $\d X$ sitting
at position $i$. Summing over Wick contractions and using the
orthogonality in (\ref{zammet}) gives the following correlation
function
\begin{align}
  \rho^I_J  \bigl ( \la 0 | a_i \bigr )   \rho^K_L  \bigl (a^\dagger_j  | 0 \ra  \bigr )
  = \sum_{\sigma\in S_n}
  \delta_{i\sigma(j)} \sigma^I_L (\sigma^{-1})^K_J
\end{align}
Here one only sums over permutations $\sigma$ which map position $j$
to $i$, enforced by the delta function.  This in turn leads to the
correlation function for $k$ oscillators ($k$ derivatives):
\begin{align}
   \rho^I_J \bigl  (  \la  0 | a_{i_1} a_{i_2}\dots a_{i_k} | \bigr )  
 \rho^K_L  \bigl(  a^\dagger_{j_1}a^\dagger_{j_2}\dots a^\dagger_{j_k}
   | 0 \ra \bigr ) = 
   \sum_{\rho\in S_k} \sum_{\sigma\in S_n} 
   \delta_{i_1 \sigma(j_{\rho(1)})}\dots  \delta_{i_k
     \sigma(j_{\rho(k)})}\sigma^I_L (\sigma^{-1})^K_J\ 
\end{align}
We wish to rewrite this in terms of the lowest weight states
$\rho^K_L  \bigl ( 
A^\dagger_{a_1} A^\dagger_{a_2}\dots A^\dagger_{a_k} | 0 \ra \bigr ) $.
The correlation function of these states is then
\begin{align}
 &    \rho^I_J \bigl ( \la 0 |   (A_{a_1} A_{a_2}\dots A_{a_k})\;
   \rho^{K }_L   \bigl ( A^\dagger_{b_1}A^\dagger_{b_2}\dots A^\dagger_{b_{k}}
   |0 \ra \bigr )\nn \\
&=\sum_{\rho\in S_{k}} \sum_{\sigma\in S_n} 
   J_{a_1}^{\sigma(j_{\rho(1)})}\dots J_{a_k}^{\sigma(j_{\rho(k)})} \; (J^\dagger)_{b_1}^{j_1} \dots (J^\dagger)_{b_{k}}^{j_{k}}\;  \sigma^I_L
   (\sigma^{-1})^K_J \nn\\
& =\sum_{\rho\in S_{k}} \sum_{\sigma\in S_n} D^H_{a_1 c_1}(\sigma)\dots
D^H_{a_k c_k}(\sigma) J_{c_1}^{j_{\rho(1)}}\dots J_{c_k}^{j_{\rho(k)}}   \; (J^\dagger)_{b_1}^{j_1} \dots (J^\dagger)_{b_{k}}^{j_{k}}\; \sigma^I_L
   (\sigma^{-1})^K_J \nn\\
& =\sum_{\rho\in S_k} \sum_{\sigma\in S_n} D^H_{a_{\rho(1)} b_1}(\sigma)\dots
D^H_{a_{\rho(k)} b_k}(\sigma) \;\sigma^I_L
   (\sigma^{-1})^K_J\ 
\end{align}
Now notice that $A_{a_1}^{\dagger}  A_{a_2}^{\dagger} \dots A_{a_k}^{\dagger} $ is totally
symmetric in the $a_i$ indices and thus lies in the totally
symmetric tensor product $\Sym ( V_H^{ \otimes k} )$.
Equivalently this is the projection in 
\begin{equation}
V_{H}^{ \otimes k } = \bigoplus_{ \L_1 , \L_2 } V_{ \L_1 }^{ (S_n)} 
 \otimes V_{ \L_2}^{ ( S_k) }  \otimes V_{ \L_1 , \L_2 } 
\end{equation}
to $ \L_2 = [k] $.  We can  consider the corresponding 
Clebsch-Gordon coefficients  to the
irreducible representation $\Lambda_1$.
\begin{equation}
C^{\Lambda_1,[k],m_{ \L_1 };\tau}_{ a_1\dots a_k} 
\end{equation}
The index $\tau $ runs through the multiplicity of the
irrep.  $ \L_1 \otimes [k] $ of $S_n \times S_k$ in $V_{H}^{ \otimes k
}$ which is the dimension of $V_{ \L , [k] } $ .  Equivalently, this
is the multiplicity of $ \L_1 $ of $S_n$ in $\Sym ( V_H^{ \otimes k}
)$.   Formulae for the multiplicities are
given in Section \ref{sec:multiplicities}. 
The   Clebsch-Gordan coefficients are explained in Appendix
Section \ref{sec:symmCG}.

The CG coefficients    have  the following properties.
Firstly 
\begin{align}\label{symmclebsch1} 
  C^{\Lambda_1,[k],m_{ \L_1 };\tau}_{ a_1\dots a_k}
  =C^{\Lambda_1,[k],m_{ \L_1 };\tau}_{ a_{\rho(1)}\dots a_{\rho(k)}} 
   \qquad \forall \rho \in S_k
\end{align}
which simply reflects the fact that the CG coefficient couples to the 
{\em{symmetric}} product $ \Sym (  V_H^{ \otimes k} ) $.
Secondly 
\begin{align}\label{symmclebsch2} 
   C^{\Lambda_1,[k],m_{ \L_1 };\tau}_{ a_1\dots a_k}
   D^H_{a_1 b_1}(\sigma)\dots D^H_{a_k
     b_k}(\sigma)=D^{\Lambda_1}_{m_{\L_1} m'_{ \L_1} }(\sigma)
   C^{\Lambda_1 , [k], m'_{ \L_1 }; \tau}_{ b_{1}\dots
     b_{k}}
\end{align}
which is derived in (\ref{symmcg2}).  
Finally orthogonality:
\begin{align}\label{orthogsc} 
  C^{\Lambda_1,[k],m_{ \L_1 };\tau}_{ a_1\dots a_k} 
  (C^\dagger)^{\Lambda'_1,[k],m'_{ \L'_1 };\tau'}_{ a_1\dots a_k} 
= \delta_{\Lambda_1 \Lambda_1'} \delta_{m_{ \L_1  }   m'_{ \L_1 }  } \delta_{\tau \tau'  }
\end{align}

Now define new operators transforming in the irreps. of $S_n$
\begin{align}\label{sl2ops}
  \bigl (  \cO^{n,k;\Lambda_1 ,  m_{ \L_1 } ; \tau} \bigr)^I_J  =
C^{\Lambda_1,[k],m_{ \L_1 };\tau}_{ a_1\dots a_k} 
 \; \rho^I_J \bigl ( 
   A^\dagger_{a_1}\dots A^\dagger_{a_k}  | 0 \ra \bigr )
\end{align}
Using the properties of the Clebsch-Gordon coefficients we  show
that the operators $\cO^{n,k;\Lambda_1 , m_{ \L_1 } ; \tau_{ \L } } $
 have the following correlation function
\begin{align}\label{finalcorsl2}
& \corr{ \bigl ( \cO^{n,k;\Lambda_1 , m_{ \L_1 } ; \tau } \bigr )^I_J
\bigl ( \cO^\dagger{}^{n',k';\Lambda'_1 , m'_{ \L_1' } ; \tau' } \bigr )^K_L
  }\nn \\
=&
   \delta^{nn'}\delta^{kk'}
    C^{\Lambda_1,[k],m_{ \L_1 };\tau}_{ a_1\dots a_k} 
     (C^\dagger)^{\Lambda'_1,[k],m'_{ \L'_1 };\tau'}_{ b_1\dots b_k}
     \nn \\
& \quad\quad\sum_{\rho \in S_k} \sum_{\sigma\in S_n} D^H_{a_{\rho(1)} b_1} (\sigma)\dots
D^H_{a_{\rho(k)} b_k}(\sigma)\; \sigma^I_L
   (\sigma^{-1})^K_J\
   \nn \\  =&
   \delta^{nn'}\delta^{kk'} k! \; C^{\Lambda_1,[k],\hat m_{ \L_1 };\tau}_{ b_1\dots b_k}  
     (C^\dagger)^{\Lambda'_1,[k],m'_{ \L'_1 };\tau'}_{
       b_1\dots b_k}
   \sum_{\sigma\in S_n} D^{\Lambda_1}_{m_{\L_1} \hat m_{\L_1} } (\sigma)\;
    \sigma^I_L
   (\sigma^{-1})^K_J
  \nn \\=&
   \delta^{nn'}\delta^{kk'}\delta^{\Lambda_1\Lambda'_1}
 \delta^{\tau \tau' } k!\;
 \sum_{\sigma\in S_n} D^{\Lambda_1}_{m_{\L_1} m'_{\L_1} }(\sigma)\; \sigma^I_L
   (\sigma^{-1})^K_J\ 
\end{align}

\subsection{Descendants}\label{sec:descendants}

We have a raising operator ${\bf L_+}$ (see equation \eqref{bfL}) corresponding to a space-time
derivative.  Acting on the lowest weight state we obtain the
descendant operator
\begin{align}
  \cO^{\L = n+k, M_\L;\Lambda_1 ,  m_{ \L_1 } ; \tau } = ({\bf L_+})^{M_\L}\;
C^{\Lambda_1,[k],m_{ \L_1 };\tau}_{ a_1\dots a_k} 
 \; A^\dagger_{a_1}\dots A^\dagger_{a_k} \ket{{\bf X}^I_J}
\end{align}
We then find the expected canonical form, using the commutator relations
in Section \ref{sec:osccon}.
\begin{align}
& \corr{ \cO^{\L = n+k, M_\L;\Lambda_1 ,  m_{ \L_1 } ; \tau }
(\cO^\dagger){}^{\L' = n'+k', M'_{\L'};\Lambda_1' ,  m'_{ \L_1' } ; \tau' }
} \nn \\
& = 
   \delta^{nn'}\delta^{kk'}\delta^{M_\L M'_{\L'}}\delta^{\Lambda_1\Lambda'_1}
 \delta^{\tau \tau' }\;(M_\L!)^2\; k!\;
 \sum_{\sigma\in S_n} D^{\Lambda_1}_{m_{\L_1} m'_{\L_1} }(\sigma)\; \sigma^I_L
   (\sigma^{-1})^K_J
\end{align}

\section{Multiplicities of $SL(2) \times S_n $ irreps. in $V_F^{\otimes n } $}
\label{sec:multiplicities}  

In this section we give formulae for  the multiplicities of the operators
found in~(\ref{sl2ops}) using characters.

\subsection{Multiplicity of $SL(2)$ irreps.} 

We begin by considering  the multiplicities of $SL(2)$ irreps in
$V_F^{ \otimes n } $ which includes a sum  over  $S_n$ irreps.  
The states $ \partial^l  X $  in $V_F$ have weights $ L_0 = 1 + l $, with 
$l$ going up to infinity. They form a lowest weight discrete series 
irrep $V_{1}=V_F $. Similar discrete series irreps exist for 
any $k$, i.e. $V_k $. 
We wish to find the tensor product decomposition of $V_1^{ \otimes n} $
in terms of the irreps. $V_k$.
This can be derived by characters. The character of the irrep. $V_k$
is  
\bea 
\chi_k(q):= Tr_{V_k }  ( q^{L_0}  ) = q^k \sum_{l=0}^{ \infty }  q^l  = { q^k \over ( 1-q) } 
\eea 
For the tensor product $V_1^{ \otimes n } $ 
we get the character
\bea 
(\chi_1(q))^n &=&   { q^{n} \over (1-q ) }   { 1 \over ( 1-q)^{n-1} } \cr
 & = &  { q^{n} \over (1-q ) } \sum_{k \ge 0 } ~~ 
{( n-2 + k )! \over k! ( n-2 )! }~~  q^k \cr
&=& \sum_{k\ge 0} \chi_{n+k}(q) ~~ m(k,n)
\eea 
where we have defined
\bea\label{mkn}  
m ( k , n )  =  { ( n-2 +k )! \over k! ( n-2)! }  
\eea 
We thus have the	 decomposition of $V_F^{\otimes n}$ as: 
\bea\label{vpndecomp}  
V_F^{ \otimes n }  = \bigoplus_{k \ge 0 }~~  m ( k , n )~~   V_{n + k } 
\eea

We have seen in the previous section that the multiplicity $ m(k,n) $ is generated by 
$k$ powers of the 
oscillators $A_a^{\dagger} $. They transform in the $H=[n-1,1]$ 
representation of $S_n$ with  dimension $(n-1) $. The $k$-oscillator states 
transform in the symmetrised tensor product of the 
hook representation which does indeed have dimension $m(k,n)$. 
The multiplicity $m(k,n)$ can be decomposed 
into irreps of $S_n$ by finding the decomposition of 
$\Sym ( V_H^{\otimes k } ) $ into irreps of $S_n$. 
We can therefore  write 
\bea 
V_{1}^{\otimes n } &=&
\bigoplus_k  V_{ \L = n + k    } \otimes \Sym ( V_{H}^{\otimes k } ) \cr 
&=& \bigoplus_k  V^{SL(2)} _{ \L = n + k    } 
\otimes V^{S_n}_{ \L_1 } \otimes V_{ \L,\L_1 }^{\Com(  SL(2) \times S_n) }
\eea  
Here the integer $k$ runs from $0$ to
infinity and   $ \Com(SL(2) \times S_n ) $ is the commutant of $SL(2)
\times S_n$.  Now if we act with a projector $P_{\L_1}  = 
{ d_{\L_1}  \over n! }
\sum_{\sigma } \chi_{\L_1}  ( \sigma ) \sigma $ on $ V_{1}^{\otimes n } $ we
will project out the subspace with  a fixed $ \L_1$. Taking the $SL(2)$ 
character
in this $\L_1$-symmetrised subspace and expanding in terms of the
characters of $ V_{ p } $ will yield the dimensions
of the multiplicity spaces $ V_{ \L,\L_1}^{\Com(  SL(2) \times S_n ) }$. 
This provides a way
to solve the problem of decomposing $\Sym ( V_{H}^{\otimes k } )$ into
irreps of $S_n$. We consider this problem in the next section.

\subsection{Multiplicities of  $SL(2)\times S_n $ and $q$-deformed $GL( \infty ) $}

\subsubsection{Examples of symmetric and antisymmetric $S_n$ irreps} 
 
As an example of this method,  take $\L_1 = [n]$
  the symmetric irrep. We want to 
calculate $tr_{\cW} P_{\L_1}  q^{L_0} $ where the trace is taken over
$\cW=V_1^{\otimes n}$. This means calculating 
$q^{L_0}$ in the symmetrised subspace of $V_1^{\otimes n}$. 
A basis in the symmetrised  subspace of $ |m_1 , m_2 , .. , m_n \ra  $
is in $1-1$ correspondence with  natural numbers 
$m_1 ,m_2 , \cdots m_n $ obeying 
\bea 
0 \le m_1 \le m_2 \le \cdots m_n \le \infty 
\eea 
So the character is 
\bea\label{symsum}  
 tr_{\cW} P_{ [n]} 
 q^{L_0} & = &  q^n  \sum_{m_n=0}^{\infty} \sum_{m_{n-1} =0}^{m_n }
\cdots  \sum_{m_2=0}^{m_3 } \sum_{m_1=0}^{m_2}  q^{m_1+ m_2 + \cdots + m_n } 
\cr 
& = &   q^{n} \prod_{i=1}^n { 1 \over ( 1 -q^i ) } \cr 
& = &   { q^{n} \over ( 1 - q ) }  \prod_{i=2}^n { 1 \over ( 1 -q^i ) }  
\eea

The multiplicity of $V_{ \L = n + k }^{SL(2) }
  \otimes V_{ [n] }^{ (S_n) }   $ is then the coefficient of $q^k$ 
in the generating function 
\bea\label{symgenfun} 
\prod_{i=2}^n { 1 \over ( 1 -q^i ) }
\eea
 As an example 
for $n=2$, the multiplicity of $ V_{ 2 + k } $ is 
the coefficient of  $q^k $ in $ { 1 \over 1-q^2 }$. This tells us that 
the symmetric irrep. of $S_n$  only appears for $k=0,2,4 , \cdots $ with unit 
multiplicity.

Similarly, for $R=[1^n]$ we  apply the antisymmetric projector to $\cW$ we have a basis 
in correspondence with $ ( m_1 , m_2 , \cdots ,m_n ) $ with 
$ m_1 < m_2 < \cdots < m_n $. So the character is 
\bea\label{asymsum}  
tr_{\cW }  ( P_{[1^n]} q^{L_0 }  )    && =
 q^n \sum_{ m_n= n-1 }^{\infty}  \sum_{ m_{n-1} = n-2 }^{ m_{n-1} - 1  } 
 \cdots \sum_{ m_2 = 1  }^{ m_3-1   } \sum_{ m_1 = 0}^{m_2-1} 
 q^{m_1 + m_2 + \cdots + m_n } \cr 
&& = q^{n} q^{n(n-1) \over 2 } \prod_{i=1}^{n}  { 1 \over 1 - q^i } \cr 
&&= { q^n \over {1-q} }  q^{n(n-1) \over 2 } \prod_{i=2}^{n} 
 { 1 \over 1 - q^i } 
\eea  
 So the  
number  of antisymmetric $[1^n]$ irreps. of $S_n$
 in the multiplicity space of $V_{n+k} $ is the coefficient of $q^k$ in    
\bea\label{asymgenfun}  
{ q^{n(n-1)\over 2 } \over  (1-q^2 ) \cdots (1-q^n ) } 
\eea 
This multiplicity is zero unless $k \ge { n(n-1)\over 2 }  $. 
This is as it should be because the antisymmetry condition 
means that we need $ X , \d X , ... \d^{n-1} X $ which 
has weight $ n + { n(n-1) \over 2 } $.

\subsubsection{The generating function for any $SL(2)\times S_n$ irreps}

In fact it turns out we can write down a compact formula for the generating function for the multiplicities of 
$ V_{ \L =  n + k } \otimes V_{ \L_1 } $ in $ \cW  $ for any $\Lambda_1 $.
It is given by 
\bea\label{conj}  
( 1-q) q^{ \sum_{i=1}  { c_i (c_i -1 )  \over 2 } } 
\prod_{ b } { 1 \over  ( 1 - q^{h_b } )  }      
\eea 
The product runs over the boxes of the Young diagram of 
$\Lambda_1$ and $h_b$ is the hook length of the box. 
$c_i$ is the column length of the $i$'th column.
 One can check that this agrees with  
(\ref{symgenfun})  and  (\ref{asymgenfun})
 for $R=[n]$ and $R=[1^n]$. 

The proof of this generating function, using $q$-dimensions of
$GL(\infty)$, goes as follows.

\subsubsection{Proof using $q$-dimensions of $GL(\infty)$} 

It is useful to think of the infinite dimensional representation 
$V_F $ as a limit of finite dimensional representations 
$V_{ \tilde N } $ of  $ GL(\tilde N )$ 
for $ \tilde N \rightarrow \infty$.  This corresponds to considering
fundamental fields of the form $\d^k X$ for $0 \leq k\leq \tilde N-1$.
States in $ V_F^{\otimes  n } = V_{ \tilde N }^{ \otimes n } $ 
of a fixed $S_n$ symmetry given by Young diagram $ \L_1 $, 
can be labelled by inserting $n$ positive integers  from 
$ 1 \cdots  \tilde N $ into the Young diagram, with the numbers 
strictly decreasing down the columns and weakly increasing 
along the rows. These are the semi-standard Young tableaux \cite{fulhar}.  
We will denote by $ \vec m ( R ) $ a set of numbers 
corresponding to a semi-standard Young tableau. This corresponds to
operators consisting of the letters
$\d^{m_1-1} X,\dots, \d^{m_n-1}X $. So we 
have 
\bea 
tr_{ \cW } P_{ \L_1 } q^{L_0} 
= \sum_{ \vec m ( R ) } q^{ m_1 + m_2 + \cdots m_n }
\eea 
In a standard basis of $ GL (\tilde N )$, with $e_i $ 
being the column vector with $1$ in the $i$'th place 
and $ 0$ elsewhere, the diagonal matrices $E_{ii} $ act as 
\bea 
E_{ii} e_m = \delta_{im} e_m 
\eea 
The weight $q^{m} $ can be identified 
with the eigenvalues of $  q^{\sum_{ m } m  E_{m m } } $. 
Precisely this generator appears in the computation of the 
$q$-dimension of the representation $\L_1$ of   $ U_q ( GL(\tilde N ) ) $. 
This sum also appears in studying the  decomposition 
of $V_j^{ \otimes n } $ for the $SU(2)$ representation $V_j$
($2j+1 = \tilde N $)   in terms of $SU(2) \times S_n $ \cite{kingpleth}. 
It is  known to be 
\bea 
 q^{n  + \sum_{i}  { c_i ( c_i -1 ) \over 2 }  } ~~ \prod_{i,j} { ( 1 - q^{\tilde N-i+j } ) \over 1 - q^{h(i,j)} }  
\eea 
where $i$ runs along the columns,  $j$ runs along the 
rows of the Young diagram, and $h(i,j)$ is the hook 
length of the box labelled by $(i,j)$. 
When $\tilde N \rightarrow \infty $ in a
region of $q < 1 $ ( where the desired sums converge), 
the numerator goes to $1$ and we get the result 
\bea 
tr_{ \cW } P_{ \L_1 } q^{L_0}  
= q^{n  + \sum_{i}  { c_i ( c_i -1 ) \over 2 }  } ~~ \prod_{i,j} { 1 \over 1 - q^{h(i,j)} } 
\eea 
Factoring out the character of $V_{\L = n+k } $ which is 
${  q^{n+k} \over ( 1 -q )  } $ we get the multiplicity of this 
representation. This proves the claim that the coefficient of 
$q^k$ in 
(\ref{conj}) is the multiplicity of the representation  $V_{\L = n+k } $
in $ \cW $.

\subsection{The general case using characters of symmetric
    groups $S_n \times S_k$ }

As we explained in Section 3.1, the oscillator construction of $SL(2)$
representations implies that the multiplicity of the $ SL(2)\times S_n
$ representation $ V_{\L = n + k } \otimes V_{\L_1} $ in the
decomposition of $V_{1}^{\otimes n } $ is given by the multiplicity of
$ V_{\L_1} $ of $S_n$ in $ \Sym ( V_H^{\otimes k } ) $.  Equivalently
this is the multiplicity of the representation $ \L_1 \otimes [k] $ of
$S_n \times S_k $ in $ V_H^{\otimes k } $, where $[k]$ denotes the
Young diagram of $S_k$ with a single row of length $k$ which is the
symmetric representation.  The projectors $ P_{\L_1} \otimes P_{[k]} $
can be written down using characters of symmetric groups.  Hence we
have
\begin{align}
  d_{\L=n+k,\L_1} & = \tr_{V_H} (P_{\L_1} \otimes P_{[k]}) \nn \\
  & = \frac{ 1 }{ n! } \sum_{ \sigma  \in S_n   } \chi_{ \L_1} ( \sigma  )\frac{1}{k!} \sum_{\tau \in S_k} \chi_{[k]}(\tau) 
    \prod_i   ( \tr_{V_H } ( \sigma^i   ))^{ c_i ( \tau ) }  
\end{align}
$c_i(\tau)$ is the number of cycles in $\tau$ of length $i$.  See
Appendix Section \ref{app:hookdecomp} for further details.

We can check that this multiplicity gives the correct $m(k,n)$ 
\begin{align}
  m(k,n) & = \sum_{\L_1(S_n)}  d_{\L_1} d_{\L = n+k,\L_1} \nn \\
  & = \sum_{ \sigma  \in S_n }\sum_{\L_1(S_n)} \frac{1}{n!}
  d_{\L_1}\chi_{ \L_1} ( \sigma  ) \frac{1}{k!}\sum_{\tau \in S_k}  \prod_i   ( \tr_{V_H } (
  \sigma^i   ))^{ c_i ( \tau ) }  \nn \\
  & = \frac{1}{k!} \sum_{\tau \in S_k}   ( \tr_{V_H } (
  \id   ))^{\sum_i c_i ( \tau ) }   \nn \\
  & = \dim_{n-1} [k] \nn \\
  & = \frac{(n-2+k)!}{k!(n-2)!}
\end{align}
We have identified $m(k,n)$ with the dimension of the totally
symmetric $GL(n-1)$ representation $[k]$.

\section{General $G, V $}\label{sec:genGV}

We now consider how the derivation of the canonical covariant 
form,  such as \eqref{finalcorsl2} in the $SL(2)$ case,  generalises
 to the situation of  any global symmetry group $G$. 
The Lie algebra generators act on $n$-fold tensor products 
of representations $V_1 \otimes V_2 \cdots \otimes V_n $ as 
\begin{equation}
  \Delta_n ( J_a ) = J_a \otimes 1\otimes \cdots\otimes 1\;\; +\;\; 1 \otimes J_a \otimes \cdots \otimes 1 
\;\;+\;\; \dots \;\;+ \;\; 1 \otimes \cdots \otimes J_a 
\end{equation}
In particular we will be interested in the $n$-fold tensor product of
the representation $V_F$ corresponding to the fundamental fields in (a
sector of) the theory. Note that the action of $G$ commutes with the
symmetric group action permuting the $n$ factors in the tensor product. 
For any  $\s \in S_n$ we have in $ End ( V_{F}^{\otimes n } ) $ 
\begin{equation}
  \sigma \;\Delta_n ( J_a ) = \Delta_n ( J_a )\;\sigma
\end{equation}
The signs that arise in the super-algebra case will be discussed 
in section \ref{signsuper}, where we will show that the key result
generalizes. 

We will now organise the states in $V_F^{\otimes n}$ according to the
representations of the product group $G \times S_n$ acting on this
space
\begin{equation}
   (V_F^G)^{\otimes n} = \oplus_{\L, \L_1} V_{\L}^G \otimes
  V_{\L_1}^{S_n} \otimes V_{\L, \L_1}^ {\Com ( G \times S_n ) }
\end{equation}

The $n$-fold tensor product has states 
\begin{equation}
  \label{basntens}  
\ket{ m_1, m_2 , \cdots , m_n }
\end{equation}
where $m$ runs over the states in the fundamental representation
(which can be infinite dimensional as in the $SL(2)$ case).  We can
choose an orthonormal basis
\begin{equation}
  \braket{ m_1 , m_2 , \cdots m_n }{ m_1' , m_2' , \cdots , m_n' } = \delta_{m_1 m_1'} \cdots \delta_{m_n m_n'} 
\end{equation}

We can decompose the $n$-fold tensor product into irreps of the global
symmetry as follows
\begin{equation}
    \label{GCG}
\ket{m_1 , \cdots , m_n } = \sum_{    \L , M_\L , i  } C_{\vec m }^{ \L , M_\L , i }\ket{    \L , M_\L , i  } 
\end{equation}
$M_\L$ is the state within $V_{\L}$ and $i$ is a multiplicity index
for  $V_{\L}$. 
In fact $ V^{\otimes n } $ has an action of $G \times S_n$ so we can
decompose the above multiplicity index $ i$ into $ ( \L_1, m_{\L_1} ,
\tau ) $ where $ \L_1 $ labels an $S_n$ irrep, $m_{\L_1}$ runs over
the states in the irrep $ \L_1 $ of $S_n$ and $ \tau $ labels the
multiplicity of $ V_{\L } \otimes V_{\L_1} $. We define the Clebsch
for this decomposition
\begin{equation}
  \label{clebdef}  
C_{\vec m }^{ \L , M_\L ,  \L_1 , m_{\L_1} , \tau   } = \braket{  \L , M_\L , \L_1 , m_{\L_1}   , \tau}{ \vec m }
\end{equation}
The Clebsch are invertible 
\begin{equation}
  \ket{  \L , M_\L , \L_1 , m_{\L_1}   , \tau} = \sum_{\vec m}C^{\vec m }_{ \L , M_\L ,  \L_1 , m_{\L_1} , \tau   } \ket{\vec m} \label{invertible}
\end{equation}
so that
\begin{equation}
  \label{clebdef2}  
C^{\vec m }_{ \L , M_\L ,  \L_1 , m_{\L_1} , \tau   } = \braket{ \vec m }{  \L , M_\L , \L_1 , m_{\L_1}   , \tau}
\end{equation}

Using the hermiticity of the 
 inner product 
\begin{equation}
  C^{\vec m }_{ \L , M_\L ,  \L_1 , m_{\L_1} , \tau   } = \left( C_{\vec m }^{ \L , M_\L ,  \L_1 , m_{\L_1} , \tau   }\right)^*
\end{equation}

We can always choose an orthonormal  basis
\begin{equation}
  \braket{ \L , M_\L ,  \L_1 , m_{\L_1} , \tau }{\L' , M'_{\L'} ,  \L_1' , m'_{\L_1'} , \tau' } = \delta_{\L\L'} \delta_{ M_\L M'_{\L'}} \delta_{\L_1\L_1'} \delta_{m_{\L_1}  m'_{\L_1'}} \delta_{\tau \tau'}
\end{equation}
which leads to orthogonality of the Clebsch

\bea\label{orthogLMI}
\fbox{
$\displaystyle{ 
 \sum_{\vec m}\left( C_{\vec m }^{ \L , M_\L ,  \L_1 , m_{\L_1} , \tau   }  \right)^*C_{\vec m }^{ \L' , M'_{\L'} ,  \L_1' , m'_{\L_1'} , \tau'  } = \delta_{\L\L'} \delta_{ M_\L M'_{\L'}} \delta_{\L_1\L_1'} \delta_{m_{\L_1}  m'_{\L_1'}} \delta_{\tau \tau'}\  
}$
}  
\eea

\subsection{Correlators in free field theory}\label{sec:corfree}

Corresponding to the orthogonal states in $ V_F $ we have 
fields $W_m$ which diagonalize the Zamolodchikov metric
\bea\label{orthogbasic}  
\la W_m W_{m'} \ra = \delta_{ m m' } 
\eea 
For  the states in $ V_F^{ \otimes n } $ take operators 
$  \cO_{m_1 , m_2 , \cdots m_n } \equiv \cO_{\vec m } = W_{m_1} \otimes W_{m_2} \cdots W_{m_n} $. 
 In the case of $ SL(2) $
this is 
\bea 
 { 1 \over m_1 ! m_2! \cdots m_n! }  
\d^{m_1} X \otimes \d^{m_2}  X \cdots \otimes \d^{m_n} X 
\eea 
The 2-point function can be written as 
\begin{equation}
 \langle  ( ~  \cO^\dagger_{m_1 , m_2 , \cdots m_n } ~ )^I_J   ,  (~   \cO_{m_1' , m_2' \cdots m_n' } ~ )^K_L  \rangle  = \sum_{ \sigma \in S_n } \prod_{i=1}^n   \delta_{ m_i m'_{\sigma(i) }  }  ~~  ( \sigma )^K_J   ( \sigma^{-1} )^I_L  
\end{equation}
where the sum over $\sigma $ runs over Wick contractions and 
we have used the orthogonality (\ref{orthogbasic}).  
Define operators in correspondence with the orthonormal $ G \times S_n$  basis 
of the $ V^{ \otimes n } $  :  
\begin{equation}\label{defop} 
\cO_{\L , M_\L ,  \L_1 , m_{\L_1} , \tau  } =
\sum_{ \vec m }  C^{\vec m }_{ \L , M_\L ,  \L_1 , m_{\L_1} , \tau   }  \cO_{ m_1 , \cdots , m_n }   
\end{equation}
The correlator of the gauge-covariant operators is 
\begin{align}
  & \langle(\cO_{\L , M_\L ,  \L_1 , m_{\L_1} , \tau  }^\dagger  )^I_J \; (\cO_{\L' , M'_{\L'} ,  \L_1' , m'_{\L_1'} , \tau' } )^K_L   \rangle \nn \\
  & = \sum_{\vec m, \vec m'}\left( C^{m_1 \cdots m_n }_{ \L , M_\L ,  \L_1 , m_{\L_1} , \tau   }  \right)^*C^{m_1'\cdots m_n' }_{ \L' , M'_{\L'} ,  \L_1' , m'_{\L_1'} , \tau'  } \sum_{ \sigma \in S_n } \prod_{i=1}^n   \delta_{ m_i m'_{\sigma(i) }  }  ~~  ( \sigma )^K_J   ( \sigma^{-1} )^I_L  
  \label{covcor} 
\end{align}
Solve the delta function 
\begin{align}\label{solvedelt} 
   &\langle(\cO_{\L , M_\L ,  \L_1 , m_{\L_1} , \tau  }^\dagger  )^I_J \;
 (\cO_{\L' , M'_{\L'} ,  \L_1' , m'_{\L_1'} , \tau' } )^K_L   \rangle \nn \\
   &= \sum_{ \sigma \in S_n }
 \sum_{\vec m}\left( C^{m_1 \cdots m_n  }_{ \L , M_\L ,  \L_1 , m_{\L_1} ,
 \tau   }  \right)^*C^{m_{\s^{-1}(1)} \cdots m_{\s^{-1}(n)}  }_{ \L' , M'_{\L'} ,  \L_1' , m'_{\L_1'} , \tau'  }  ~~  ( \sigma )^K_J   ( \sigma^{-1} )^I_L  
\end{align}

We can simplify the $\s^{-1}$ action on $V_F^{\otimes n}$ in the
second Clebsch because we know it transforms under the $S_n$
representation $\L_1'$: the action is just the matrix representation
of $\L_1'$.
\begin{equation}\label{clebschprop} 
  C^{m_{\s^{-1}(1)} \cdots m_{\s^{-1}(n)}  }_{ \L' , M'_{\L'} ,  \L_1' , m'_{\L_1'} , \tau'  } = D^{\L_1'}_{m'_{\L_1'} \hat{m}_{\L_1'}}(\s^{-1})\;\;  C^{m_1 \cdots m_n   }_{ \L' , M'_{\L'} ,  \L_1' , \hat{m}_{\L_1'} , \tau'  }
\end{equation}

Use this and the orthogonality of the Clebsch from equation
\eqref{orthogLMI} to get
\begin{equation}\label{finalorh} 
\fbox{ 
$\displaystyle{ 
   \langle(\cO_{\L , M_\L ,  \L_1 , m_{\L_1} , \tau  }^\dagger  )^I_J \; 
(\cO_{\L' , M'_{\L'} ,  \L_1' , m'_{\L_1'} , \tau' } )^K_L   \rangle   \\
= \delta_{\L\L'} \delta_{ M_\L M'_{\L'}} \delta_{\L_1\L_1'}
  \delta_{\tau \tau'}  \sum_{ \s \in S_n } D^{\L_1}_{m_{\L_1} m'_{\L_1'}  }
 ( \sigma )  \;   (
     \sigma )_J^{K}  (\s^{-1} )^I_L
}$}
\end{equation}

\subsubsection{Signs in the super-algebra case }\label{signsuper} 
In the case where the generators of $G$ include 
fermionic ones, there is a small modification of the 
above proof. Fermionic generators $Q $ pick up signs 
when taken past fermionic fields
\bea 
Q ( \psi_1 \psi_2 ) = ( Q \psi_1 ) \psi_2 - \psi_1 ( Q \psi_2 ) 
\eea 
In this case the action of permutations is defined 
 with a sign for each fermion swap. For the transposition of 
$ \psi_1  \psi_2 $ we define 
\bea 
s  ( \psi_1 \psi_2 ) =  - \psi_2 \psi_1 
\eea 
It is easy to check that $ Qs = s Q $. 
This is the key point. If we define the action of permutations
to pick up a sign for every swap of fermions, we have an 
action of the permutation group which commutes with 
the super-algebra.  Hence Clebsch-Gordan coefficients for 
$ G \times S_n$ are well-defined, and we can define 
operators according to (\ref{defop}). The two-point function in 
this case picks up a sign $ (-1)^{ \epsilon ( \vec m  , \sigma )} $
on the RHS of (\ref{covcor}).   This sign carries into 
(\ref{solvedelt}). Because the correct action of the permutations 
involves this same sign factor,  equation (\ref{clebschprop}) has the sign 
$ (-1)^{ \epsilon ( \vec m  , \sigma ) }$ on the left. 
The final result (\ref{finalorh}) remains unchanged.

\subsection{Completeness}

The completeness of these operators in $V_F^{\otimes n}$ follows from
the invertibility of the Clebschs; this means that any state in
$V_F^{\otimes n}$ can be written as a linear combination of states in
$V_\L \otimes V_{\L_1}$.
\begin{equation}
 \cO_{ m_1 , \cdots  m_n }  = \sum_{\L , M_\L ,  \L_1 , m_{\L_1} , \tau  }C_{\vec m }^{ \L , M_\L ,  \L_1 , m_{\L_1} , \tau   }  \cO_{\L , M_\L ,  \L_1 , m_{\L_1} , \tau  }
\end{equation}

If we reintroduce the gauge indices and trace our covariant operators
to get gauge-invariant operators, then we must also take into account
finite $N$ constraints.  This is studied in Section \ref{sec:gaugeN}.

\section{Examples and applications} 

The above formalism can be applied to any global symmetry $G$ of a
theory, or to a subgroup of the global symmetry acting on a sector of
the fields.  In this section we will consider a number of examples to
which we can apply this formalism.  Examples relating to $\cN$=4 SYM
are the $SL(2)$ sector described above and the $U(M)$ (BPS) sector
considered in~\cite{bhr}.  We will review both of these sectors from
the new perspective of the previous section.
Free conformal field theories also
possess an enlarged symmetry group known as higher spin symmetry
and we will also see in this section  how our
construction naturally assembles into representations of this. 

Another sector we will consider corresponds to the $6$ 
hermitian scalars of $ \cN = 4 $ SYM and the $SO(6)$ subgroup 
of the global symmetry acting on them. The problem 
of diagonalising the gauge invariant  operators 
amounts to first finding a manageable form of the 
 Clebsch-Gordan coefficients for the $SO(6) \times S_n$ decomposition of 
the $n$-fold tensor product of the fundamental of $SO(6)$. 
We will not solve this problem explicitly but will
note some facts about the relevant multiplicities in Section 
\ref{sec:so6}. A more explicit description of the Clebsch-Gordans
analogous to what we gave for $U(M)$ in Section \ref{sec:um}
would be desirable.

The sector of six scalars can also be described 
in terms of the $3$ complex scalars and their conjugates. 
In this case, it is natural to use the $U(3)$ subgroup 
of $SO(6)$. The case of purely holomorphic operators 
was solved in \cite{bhr} and is reviewed in \ref{sec:um}.
It should be possible to  include the anti-holomorphic operators 
by using $U(3)$ along with the Brauer algebra $B_N ( m,n ) $
used in \cite{kimram} for the case of a single complex scalar.

Finally we will look at the formalism applied to product groups.  To
handle a class of derivatives of the holomorphic scalars we can use
the product group technology applied to $ SL(2) \times U(3)$. A simple
sector including one chiral fermion, 3 holomorphic scalars is
controlled by $ SL(2 ) \times U (3|1) $.  For the six hermitian
scalars and a class of their derivatives we can use $ SL(2 ) \times
SO(6)$. For more general derivatives we can use $ SO(4,2) \times
SO(6)$. For the sixteenth BPS sector we would use $ SU(3|2,1)$ 
\cite{Kinney:2005ej}.For 
the complete set of fields of $\cN=4 $ SYM we can use the
full symmetry $ SU( 2,2 |4)$. Calculating the Clebsch-Gordan
coefficients for the $ V_F^{\otimes n } $ will allow, following the
derivation of (\ref{finalorh}) from (\ref{orthogLMI}), to get the
canonical covariant form which in turn leads to diagonal gauge
invariant operators using Section \ref{sec:gaugeN}.  The relevant
Clebsch-multiplicities are known for $n=2$~\cite{beisert}.  Finding the
multiplicities and the Clebsch-coefficients in terms of symmetric
groups, as we do below for $ U(M ) $ and $SL(2)$, is the next step in
the solution of the free field diagonalisation problem for $ \cN =4 $
SYM.

\subsection{$ SL(2) $ and $ \Sym ( V_H^{\otimes k} ) $} 

Here we illustrate our scheme with the construction of the Clebsch
$C^{\vec m }_{ \L , M_\L , \L_1 , m_{\L_1} , \tau }$ for the specific
example of $G = SL(2)$, which we considered in section \ref{sec:SL2}. 

In the Fock space of oscillators we have
\begin{align}
  \ket{m_1 , \cdots , m_n } \sim  (a_1^\dagger)^{m_1} \dots
    (a_n^\dagger)^{m_n} | 0 \ra 
\end{align}
whereas the oscillators $A_a$ give the decomposition in terms of
$SL(2)$ irreps $\L = n+k$, following equation~\eqref{GCG}
\begin{align}
   \ket{   \L, M ; (a_1,
  \dots a_k) } =  C_{\vec{m}}^{\L,M,(a_1 \dots a_k)} \ket{m_1 , \cdots , m_n} = A^\dagger_{(a_1} \dots A^\dagger_{a_k)} 
( {\bf L}_{+}  )^M | 0 \ra
\end{align}
so the label $i$ in equation~(\ref{GCG}) is given by the indices
$(a_1, \dots a_k) $. The Clebsch-Gordon coefficients of
equation~(\ref{GCG}) can now be read off.

The next step is to decompose the label $i$ into irreps of $S_n$. The
indices $a_i$ carry the $n-1$ dimensional hook representation, $H=
[n-1,1]$ of $S_n$. Therefore the label $i=(a_1 \dots a_k)$ carries the
reducible representation $\Sym V_H^{\otimes k}$. This decomposes
into the irreducible representations $\Lambda_1$ of $S_n$ with
multiplicity $\tau$ via the Clebsch Gordon coefficient
$C^{\L_1,m_{\L_1},\tau}_{(a_1 \dots a_k)}$
\begin{align}\label{sncg}
\ket{(a_1 \dots a_k)}=  C_{\L_1,m_{\L_1},\tau}^{(a_1 \dots a_k)} \ket{\L_1,m_{\L_1},\tau}
\end{align}
and the inverse transformation is given by 
$C_{\L_1,m_{\L_1},\tau}^{(a_1 \dots a_k)}$ as 
\bea 
\ket{\L_1,m_{\L_1},\tau} =  C^{\L_1,m_{\L_1},\tau}_{(a_1 \dots a_k)}
\ket{(a_1 \dots a_k)}
\eea 
Putting all this together we get
\begin{align}
 \ket{\L,M,\L_1,m_{\L_1},\tau} & =  C^{\L_1,m_{\L_1},\tau}_{(a_1 \dots a_k)} \ket{   \L, M ; (a_1,
  \dots a_n) }\nn \\
& = C^{\L_1,m_{\L_1},\tau}_{(a_1 \dots a_k)}  C_{\vec{m}}^{\L,M,(a_1 \dots a_k)} \ket{m_1 , \cdots , m_n}
\end{align}
The Clebsch of equation \eqref{invertible} is given by
\begin{equation}
  C_{\vec m }^{ \L , M , \L_1, m_{\L_1} , \tau }= 
C^{\L_1,m_{\L_1},\tau}_{(a_1 \dots a_k)}  C_{\vec{m}}^{\L,M,(a_1 \dots a_k)}
\end{equation}

\subsection{$U(M)$ revisited}\label{sec:um}

This formalism also applies to the $U(M)$ case studied
in~\cite{bhr}. In that case the starting point is
\begin{align}
  \ket{m_1 , \cdots , m_n } \sim X_{m_1} \otimes \cdots \otimes 
 X_{m_n}\qquad m_i=1\dots M
\end{align}
which is the tensor product of $n$ fundamentals of $U(M)$,
$V_M^{\otimes n}$.  Then in~(\ref{GCG}) $C_{\vec m }^{ \L , M , i }$
is the $U(M)$ Clebsch Gordon coefficient decomposing $V_M^{\otimes n}$
into irreps $\Lambda$ with multiplicity $i$. In this case Schur Weyl
duality tells us that $i$ also carries the fundamental of the
representation $\L$ of $S_n$.  Thus $\L_1 = \L$ and there is no $\tau$
multiplicity.

To be explicit, to get the operators of \cite{bhr}
\begin{equation}\label{bhrcg}
  C_{\L,M,\L_1 =\L,m_{\L_1} = i}^{\vec{m}} = \frac{1}{n!}
  \sum_{\s \in S_n} B_{j\b} D_{ij}^{\L} (\s) \prod_{k=1}^{n}
  \delta_{m_k p_{\s^{-1}(k)}}
\end{equation}
Here $M = (\m,\b)$. $\mu$ labels the number of different flavour
fields in the operator ($\m_1$ $X$'s, $\m_2$ $Y$'s, etc.), while $\b$
runs over the number times the trivial representation of $H_\m =
S_{\m_1} \times \cdots S_{\m_M}$ is contained in $S_n$.  $B_{j\b}$ is
a branching coefficient for the change of basis for the subspace of
the irrep. $\L$ invariant under $H_\m$. Canonically we choose $p_1,
\dots p_{\m_1} = 1$, $p_{\m_1 +1}, \dots p_{\m_1 +\m_2} = 2$, \dots.
With this choice we recover the covariant operators in \cite{bhr}
\begin{equation}\label{umcovops}
 \cO^{\L\m}_{i\b} = \sum_{\vec{m}}C_{\L,\m,\b,i}^{\vec{m}} \;X_{m_1} \dots X_{m_n} = \frac{1}{n!}
  \sum_{\s \in S_n} B_{j\b} D_{ij}^{\L} (\s) \;\s {\bf X}^{\m} \s^{-1}
\end{equation}

We then find the orthogonality we expect, up to a normalisation factor
\begin{equation}
  \sum_{\vec{m}}C_{\L,\m,\b,i}^{\vec{m}} C_{\L',\m',\b',i'}^{\vec{m}}
  = \delta_{\L\L'}\delta_{\m\m'}\delta_{\b\b'} \delta_{ii'}
  \frac{|H_\m|}{n!d_{\L}} 
\end{equation}

\subsection{The higher spin group}

 The free theory of $\cN$=4 SYM is invariant under an infinite
 dimensional group $HS(2,2|4)$ known as the higher spin group. In the
 interacting theory this is
 broken to the superconformal group $SU(2,2|4)$ but it can
 nevertheless be useful
 for some applications (eg possible relations via AdS/CFT to a possible
 `tensionless limit' of string theory) to consider this enlarged
 group. When restricted to the $SL(2)$ sector the higher spin group is
 known as $HS(1,1)$. Operators form lowest weight representations of
 $HS(1,1)$ (which further decompose into an infinite number of lowest
 weight representations of $SL(2)$.) The lowest weight states of these
 representations were decscribed in\cite{bianchi}.
 In terms of the oscillators introduced in
 section~\ref{sec:osccon}, the higher spin algebra is spanned  by the
 generators
 \begin{align}
   J_{p,q}=\sum_i (a_i^\dagger)^p (a_i)^q
 \end{align}
which clearly contains the $SL(2)$ algebra~(\ref{bfL}).

If we consider truncating the fundamental fields so that we only
consider states $\ket{m}=(a^\dagger)^m\ket0$ for $m\leq M-1$ then the
higher spin group truncates to $U(M)$. Therefore the covariant
canonical form corresponding to the  higher spin group is simply the
$M\rightarrow \infty $ limit of that in the previous subsection.

Therefore the results of the previous subsection generalise naturally to
the higher spin case. Irreducible representations of the higher spin
group are specified 
by Young Tableaux, $\Lambda_1$, (as observed in~\cite{bianchi}).
We have
\begin{align}
  V_F^{ \otimes n } &= \bigoplus_{ \L_1 \vdash n }~~  V_{ \L_1 }^{ HS } \otimes
  V_{\L_1}^{ S_n }\\
&=\bigoplus_{  \L_1 , \L }  V_{\L  }^{ SL(2) }  
 \otimes V_{ \L , \L_1 }^{ \Com ( SL(2) \times S_n ) } \otimes V_{\L_1}^{ ( S_n ) } 
\end{align}
The first line is the standard Schur-Weyl duality for $U(M)$ in the
limit $M\rightarrow \infty$. Each  higher spin representation,
$\Lambda_1$, then decomposes further into an $SL(2)$ irrep $\Lambda$
and the commutant.
The Clebsch gving the first line is given by~(\ref{bhrcg}). The canonical
covariant operators which form irreps of the higher spin
group are given by~(\ref{umcovops}) with $X_m$ replaced by $(1/m!)\partial^m X$.

\subsection{The $SO(6)$ sector}\label{sec:so6}

We have $6$ hermitian scalar matrices in 
$\cN=4$ SYM, transforming in the fundamental of 
$SO(6)$. We know from the general discussion 
in Section \ref{sec:genGV} that the $SO(6)$ covariant 
diagonalisation of free field correlators will be solved 
once we have solved the Clebsch-Gordan problem for 
$SO(6) \times S_n $ in $V^{ \otimes  n } $. Here $V$ is the 
fundamental of $SO(6)$. 
\bea\label{so6decomp}  
  V^{ \otimes n } && =  \bigoplus_{ \L_1  } V_{    \L_1 }^{ ( S_n )} \otimes 
        V^{ GL(6)}_{ \L_1  } \cr 
&&= \bigoplus_{  \L_1 , \L_2 }  V_{\pi (  \L_2 ) }^{  ( SO(6) ) }  
 \otimes V_{ \L_1 , \L_2 } \otimes V_{\L_1}^{ ( S_n ) } 
\eea    
We first decompose the $n$-fold tensor space according to 
the $S_n$ symmetry. The Schur-Weyl dual of $S_n$ is $GL(6)$ 
hence the decomposition in the first line. In the second line, 
we decompose the $GL(6)$ representations to $SO(6)$ representations. 
The dimension of the multiplicity space $ V_{ \L_1 , \L_2 } $ 
is given by 
\bea 
Dim V_{ \L_1 , \L_2 }  = \sum_{ \delta } g ( \L_2 ,  2 \delta ; \L_1  ) 
\eea  
$\L_1$ is a Young diagram with $n$ boxes, $ 2 \delta $ is a partition
with even parts, i.e. Young diagram with even row lengths. The sum above
 includes   a sum over $k \ge 0 $, where $2k$ is the number of 
boxes in $ 2 \delta $ and $ n-2k$ is the number of boxes in $\L_2$.

The representations of $GL(6)$ are labelled by 
Young diagrams with  row lengths $ \l_1 \ge  \l_2 \ge  \cdots  \l_6  \ge 0 $. 
The representations of $SO(6)$ are labelled by 
$ \l_1 \ge \l_2 \ge  | \l_3 | \ge 0  $.   The last label 
$ \l_3 $ can be positive or negative. For $ \l_3 =0 $ the 
irreps are constructed by symmetrising according to the Young 
diagram and projecting out traces. When $|\l_3| > 0 $ 
the corresponding operation of Young-symmetrising and  
removing traces leaves us with a reducible representation, 
which is a direct sum of irreps. $ ( \l_1 , \l_2 , \l_3 ) \oplus
 ( \l_1 , \l_2 , - \l_3 )$. The operation $\pi $ which appears 
in (\ref{so6decomp}), when it  acts on any $GL(6)$ Young diagram $\L_1$ 
gives either zero or a Young diagram obeying the $SO(6)$ constraints. 
It is defined in terms of an operation on Young diagrams in \cite{koiter}.  

We have arrived above at the $SO(6) \times S_n$ decomposition 
by first decomposing into $ GL(6) \times S_n $, then reducing the 
$GL(6)$ to $SO(6)$. We can equally start by decomposing  in terms of 
$SO(6) \times E_6(n)$ where $E_6(n)$ is the commutant of 
$SO(6) $ in $V^{ \otimes n } $ described for example in \cite{grood}. 
A subsequent decomposition of $E_6(n)$ to $S_n$ should yield 
the same result as (\ref{so6decomp}). This follows from general 
theorems on double commutants which assert that if    $A$ is a subalgebra 
of $B$, and  $End (  B ) \subset End (A)$ are 
 their commutants in some vector 
space, then   the reduction multiplicities for irreps of  
$ B \rightarrow A $ coincide with those of  $ End ( A ) \rightarrow End( B ) $
(see \cite{halverson}  ). In this case the reduction multiplicities
of $ GL(6) \rightarrow SO(6) $ coincide  with those of
  $E_6(n) \rightarrow S_n$.

\subsection{The $SO(4,2)$ sector}\label{sec:SO42} 

In considering the sector of a scalar field $X$ with all four derivatives 
acting on it, we can  use the $SO(4,2)$ symmetry. 
Generalizing the linear combinations  $ A^{\dagger}_{ a } $ 
of oscillators which generate the lowest weights in the $SL(2)$ sector, 
we now have $ A^{\dagger}_{ a \l } $ where $\l$ is an index in 
the fundamental of $ SO(4) \subset SO(4,2)$ and as before $a$ 
is in the hook representation $V_H =  [n-1,1]$ of $S_n$.
Lowest  weights annihilated by $ K_{\l } $, with $k$ 
derivatives acting on $n$-field composites  can be constructed from
 oscillators 
of the  form 
\bea\label{so4form}  
A^{\dagger }_{ a_1 \l_1  }  A^{\dagger }_{ a_2 \l_2  }
\cdots  A^{\dagger }_{ a_k \l_k  } | 0 \ra 
\eea 
The simplest class of such LWS are those in which
 the indices  $ ( \l_1 , \l_2 , \cdots , \l_k )$ 
are taken to be a  symmetric traceless $SO(4)$ tensor corresponding to 
the $ SO(4)$ Young diagram $ [k ] $. 
These states satisfy a  type of extremality condition $ L_0 = n + k $. 
More generally we will have states of the form (\ref{so4form}) 
which involve contractions of the $ \l_i $. In these cases 
we have to mod out by the equations of motion, which leads to 
a projection of  the $ Sym ( V_H \otimes V_H ) $ representation 
of $ A^{\dagger}_{ a_1 \l }  A^{\dagger}_{ a_2 \l } $ 
 to the $S_n$ representation $[n-2,2]$. This has dimension 
${ n(n-3)\over 2 } $ which is the number obtained by  subtraction 
of $n$, for the equations of motion, from the dimension 
$ { n ( n-1 ) \over 2 } $ of $Sym ( V_H \otimes V_H ) $.  
Work on a complete solution of the diagonalisation 
in this sector, using the above facts to give a symmetric group 
description of the $ SO(4,2) \times S_n $ Clebsch-Gordans, is in progress. 
It is clear that the symmetric $SO(4)$ operators
involving the contractions will have $ L_0 > n + k $.  The 
`extremal' operators mentioned above will be useful in the 
comparison to excitations of half-BPS giants in Section \ref{sec:excite}.

\subsection{Fields carrying reps of product groups} 

  Suppose the global
symmetry group has the form $G_1 \times G_2 $.  We consider a field
$\Psi_{k,m }$ where $k$ is an index transforming under irrep $V_1$ of
$G_1 $ and $m $ transforms under irrep $V_2$ of $G_2$. Consider the
covariant operator
\begin{equation}
   ( \cO_{k_1, m_1 ; k_2 ,m_2 ; \cdots ; k_n , m_n }  )^{ I }_{J}  \equiv    (\Psi_{k_1,m_1 })^{i_1}_{j_1}
  (\Psi_{k_2,m_2})^{i_2}_{j_2}    \cdots  (\Psi_{k_n,m_n })^{i_n}_{j_n}   
\end{equation}
Fields with $n$ factors transform under the irrep $ ( V_1 \otimes V_2
)^{\otimes n} $.  With $ \sigma \in S_n $ acting simultaneously on $V_1
$ and $V_2$, the commutant of $G_1 \times G_2$ contains $S_n$. We can
consider the group $G_1 \times G_2 \times S_n $ acting on the
$n$-field composites. Correspondingly  there is a  decomposition of
the $n$-fold tensor product into irreps. of $ G_1 \times G_2 \times
S_n $.  The irreps are related to the product states as
\begin{equation}
  \ket{ \L_1, M_{\L_1 }  , \L_2, M_{\L_2} , \L_3  , m_{\L_3} , \tau }
= C_{ \L_1 , M_{\L_1 } , \L_2, M_{\L_2}  , \L_3 , m_{\L_3} , \tau }^{ \vec k, \vec m  } \ket{ \vec k , \vec m } 
\end{equation}
$\L_1$ is an irrep of $G_1$, $\L_2$ of $G_2$ and $\L_3$ of $S_n$.
Conversely
\begin{equation}
  \ket{ \vec k, \vec m  } =  C_{ \vec k,\vec m  }^{  \L_1, M_{\L_1 }  , \L_2 , M_{\L_2}, \L_3  , m_{\L_3} , \tau } \ket{\L_1, M_{\L_1 } , \L_2 , M_{\L_2} , \L_3  , m_{\L_3} , \tau}
\end{equation}
In terms of vector spaces this decomposition is
\begin{equation}
   ( V_1 \otimes V_2 )^{\otimes n} = \bigoplus_{\L_1,\L_2,\L_3} V_{\L_1}^{G_1}\otimes V_{\L_2}^{G_2}\otimes V_{\L_3}^{S_n}\otimes V_{\L_1,\L_2,\L_3}^{\Com(G_1\times G_2\times S_n)} \label{prodcanon}
\end{equation}
$\tau$ over the multiplicity space $V_{\L_1,\L_2,\L_3}^{\Com(G_1\times
  G_2\times S_n)}$.

The orthogonality of  Clebsch-Gordan coefficients is the same as before
%\begin{align}
%&   \sum_{ \vec m , \vec k } \left(C_{\vec m , \vec k }^{  \L_1 , \L_2 , \L_3 , M_{\L_1 } , M_{\L_2} , m_{\L_3} , \tau }\right)^*C_{\vec m , \vec k }^{  \L_1' , \L_2' , \L_3' , M'_{\L_1' } , M'_{\L_2'} , m'_{\L_3'} , \tau'} \nn \\
%& = \delta_{\L_1\L_1'}\delta_{\L_2\L_2'}\delta_{\L_3\L_3'} \delta_{M_{\L_1}  M'_{\L_1'}}  \delta_{M_{\L_1}  M'_{\L_1'}}  \delta_{m_{\L_3}  m'_{\L_3'}} \delta_{\tau \tau'}
%\end{align}
and if we define operators 
\begin{equation}
  (\cO_{   \L_1 , M_{\L_1 }, \L_2  , M_{\L_2} , \L_3 , m_{\L_3} , \tau  } )^I_J  = \sum_{ \vec k , \vec m  }  C^{ \vec k,\vec m  }_{  \L_1  , M_{\L_1 }, \L_2, M_{\L_2} , \L_3  , m_{\L_3} , \tau }(\cO_{\vec k,\vec m  })^I_J
\end{equation}
then the covariant two-point function is diagonal in the $( \L_1
,M_{\L_1 } , \L_2 , M_{\L_2} , \L_3 , \tau )$ indices, exactly
analogous to the single group case in equation \eqref{finalorh}.

\subsubsection{Product Clebsch in terms of single group Clebschs}

Another way that we could organise $ ( V_1 \otimes V_2 )^{\otimes n}
$, in contrast to the $G_1\times G_2 \times S_n$ decomposition in
\eqref{prodcanon}, is in terms of the separate groups
\begin{align}
  ( V_1 \otimes V_2 )^{\otimes n} & =  V_1^{\otimes n} \otimes V_2^{\otimes n} \nn \\
  & = \left(\bigoplus_{\L_1,\L_4} V_{\L_1}^{G_1}\otimes V_{\L_4}^{S_n} \otimes V_{\L_1,\L_4}^{\Com(G_1\times S_n)} \right)\otimes \left(\bigoplus_{\L_2,\L_5} V_{\L_2}^{G_2}\otimes V_{\L_5}^{S_n} \otimes V_{\L_2,\L_5}^{\Com(G_2\times S_n)}  \right) \nn
\end{align}
We use the Clebsch $ C_{ \vec k }^{ \L_1, M_{\L_1 } , \L_4 , m_{\L_4}
  , \tau_1}$ for $G_1$ and $ C_{ \vec m }^{ \L_2 , M_{\L_2 },\L_5 ,
  m_{\L_5} , \tau_2}$ for $G_2$.  Given the simultaneous action of
$S_n$ on $ ( V_1 \otimes V_2 )^{\otimes n} $, to connect this
decomposition with that in \eqref{prodcanon} we tensor together the
two $S_n$ irreps $V_{\L_4}^{S_n}$ and $V_{\L_5}^{S_n}$ to get the
irrep of the simultaneous $S_n$ action $V_{\L_3}^{S_n}$
\begin{equation}
  V_{\L_4}^{S_n} \otimes V_{\L_5}^{S_n} = \bigoplus_{\L_3} V_{\L_3}^{S_n}\; C(\L_4,\L_5;\L_3)\label{clebschsn}
\end{equation}
$C(\L_4,\L_5;\L_3)$ counts the number of times $V_{\L_3}^{S_n}$
appears in the $S_n$ tensor product $V_{\L_4}^{S_n} \otimes
V_{\L_5}^{S_n}$.  This construction shows us how to write down the
relation between the $G_1\times G_2 \times S_n$ Clebsch and the $(G_1
\times S_n)\times (G_2 \times S_n) $ Clebschs
\begin{equation}
   C_{\vec k , \vec m  }^{  \L_1 ,M_{\L_1 } , \L_2 , M_{\L_2} , \L_3 ,  m_{\L_3} , \tau } =  C_{ \vec k }^{ \L_1,M_{\L_1 } ,  \L_4 , m_{\L_4} ,
  \tau_1} \;C_{ \vec m }^{ \L_2 ,M_{\L_2 } ,\L_5 , 
  m_{\L_5} , \tau_2}  \; C^{\tau_3}{}^{\L_3}_{ m_{\L_3}}\;{}^{\L_4}_{ m_{\L_4}}\;{}^{\L_5}_{ m_{\L_5}}
\end{equation}
The $S_n$ Clebsch-Gordan coefficient $C^{\tau_3}{}^{\L_3}_{
  m_{\L_3}}\;{}^{\L_4}_{ m_{\L_4}}\;{}^{\L_5}_{ m_{\L_5}}$ gives the 
change of basis 
 for the decomposition in \eqref{clebschsn}; it
maps the states of the reps in $V_{\L_4}^{S_n} \otimes V_{\L_5}^{S_n}$
to those in $V_{\L_3}^{S_n}$.  $\tau_3$ labels the
$C(\L_4,\L_5;\L_3)$ degeneracy.  The $\tau$ which labels the product
group commutant $V_{\L_1,\L_2,\L_3}^{\Com(G_1\times G_2\times S_n)}$
is now a combination of the separate group multiplicities and the
$S_n$ tensor label $\tau_3$: $\tau = (\tau_1,\tau_2, \tau_3)$.
\begin{equation}
  V_{\L_1,\L_2,\L_3}^{\Com(G_1\times G_2\times S_n)} = \bigoplus_{\L_4,\L_5}  V_{\L_1,\L_4}^{\Com(G_1\times S_n)} \otimes V_{\L_2,\L_5}^{\Com(G_2\times S_n)} \;C(\L_4,\L_5;\L_3)
\end{equation}

In the special case when the gauge group is $U(1)$, so that the fields
commute, $ \L_3 ( S_n ) $ is the trivial representation.  This forces
$ \L_4 ( S_n ) = \L_5 ( S_n ) $. The same thing applies when
considering bosonic oscillators carrying indices of $G_1 \times G_2 $.

\section{Gauge invariant operators}\label{sec:gaugeN}
 
We have organised $n$ copies of the  fundamental fields in terms of
representations of the global symmetry group $G$.  
\begin{align}
  \cO^{\L,M_\L, \L_1, m_{\L_1}, \tau} =
  C^{\vec{m}}_{\L,M_\L,\L_1,m_{\L_1},\tau} W_{m_1} \otimes W_{m_2}
  \otimes \cdots \otimes  W_{m_n}
\end{align}

We now introduce the $U(N)$ gauge group  indices
\begin{equation}
  \left(W_{m} \right)^i_j
\end{equation}
and view the $W_{m}$ as operators on the fundamental representation 
$V_N$ of $U(N)$. 
To form gauge-invariant operators we multiply these matrices together
and then take products of traces organised by the symmetric group
element $\a \in S_n$
\begin{equation}
  \tr  ( \a \;  W_{m_1} \otimes  W_{m_2} \otimes 
   \cdots  \otimes W_{m_n}) =  (W_{m_1})^{i_1}_{i_{\a(1)}}(W_{m_2})^{i_2}_{i_{\a(2)}} \cdots (W_{m_n})^{i_n}_{i_{\a(n)}} \label{basicgio}
\end{equation}
where the trace is being taken in $V_N^{\otimes n } $. 
We can reorganise these in terms of representations $R$ of $U(N)$
using the Schur-Weyl dual $S_n$ representation matrices $D^R_{ij}
(\a)$
\begin{equation}
 \frac{1}{n!} \sum_{\a\in S_n} D^R_{ij} (\a)  \tr  ( \a \;  W_{m_1} \otimes 
 W_{m_2}
   \cdots  \otimes W_{m_n})
\end{equation}
As a representation of $U(N)$, $R$ has at most $N$ rows.

Combining the free $S_n$ indices with an $S_n$
Clebsch-Gordan coefficient gives  a gauge invariant operator
\begin{align}
  \cO^{\L,M_\L, \L_1, \tau,R,\tau_{\L_1,R}} &=
   C^{ \tau_{\L_1,R}}\;{}^{ \L_1}_{ m_{\L_1}}\;{}^R_i\;{}^R_j\;\;  C^{\vec{m}}_{\L,M_\L,\L_1,m_{\L_1},\tau} \frac{1}{n!} \sum_{\a\in S_n} D^R_{ij} (\a) 
 \tr  ( \a \;  W_{m_1} \otimes  
   \cdots \otimes  W_{m_n}) \nn \\
 \label{eq:finiteNop}& =  C^{ \tau_{\L_1,R}}\;{}^{ \L_1}_{ m_{\L_1}}\;{}^R_i\;{}^R_j ~~  \frac{1}{n!} \sum_{\a\in S_n} D^R_{ij} (\a)  \tr  \left( \a \; 
\cO^{\L,M_\L, \L_1, m_{\L_1}, \tau}\right)
\end{align}
We can invert the Clebschs to recover from these operators the basic
gauge invariant operators in \eqref{basicgio}.  This means that our
new basis is complete.  It also counts correctly at finite $N$, as
demonstrated in the next section.  Furthermore, following the methods
of \cite{bhr}, it is fully diagonal in all its labels.
\begin{align}
 & \corr{\cO^{\L,M_\L, \L_1, \tau,R,\tau_{\L_1,R}}\; (\cO^\dagger)^{\L',M'_{\L'}, \L'_1, \tau',R',\tau'_{\L_1',R'}}} \nn \\
  & =\delta_{\L\L'} \delta_{ M_\L M'_{\L'}} \delta_{\L_1\L_1'}  \delta_{\tau \tau'} 
\;\;  C^{ \tau_{\L_1,R}}\;{}^{ \L_1}_{ m_{\L_1}}\;{}^R_i\;{}^R_j\;\;C^{ \tau'_{\L_1',R'}}\;{}^{ \L_1'}_{ m'_{\L_1'}}\;{}^{R'}_k\;{}^{R'}_l\;\; \nn\\
& \frac{1}{n!} \sum_{\a\in S_n} D^R_{ij} (\a)  \frac{1}{n!} \sum_{\a'\in S_n} D^{R'}_{kl} (\a') \sum_{ \s \in S_n } D^{\L_1}_{m_{\L_1} m'_{\L_1'}  } ( \sigma )
\tr(\a \s \a' \s^{-1}) \nn \\
& =\delta_{\L\L'} \;\delta_{ M_\L M'_{\L'}}\; \delta_{\L_1\L_1'}\;  \delta_{\tau \tau'}\; \delta_{RR'} \;\delta_{\tau_{\L_1,R}\;\tau'_{\L_1',R'} }\; \frac{ n!  d_{\L_1}}{d_R^2} \Dim R
\end{align}
In the second line we have used the canonical covariant correlator
\eqref{finalorh}.  $\Dim R$ is the $U(N)$ dimension of $R$.

\subsection{Finite $N$ counting}\label{sec:countfN} 

We show here that the operators defined in equation
\eqref{eq:finiteNop} count correctly for finite $N$.  The finite $N$
partition function is given in terms of the single letter partition
function $f({\bf x})$, for bosonic ${\bf x}$, by an integral over the
$U(N)$ matrix $U$ \cite{Aharony:2003sx,Kinney:2005ej}.
\begin{equation}
   \cZ = \int [dU]
\, \exp\left\{ \sum {1 \over m} f({\bf x}^m) {\rm
tr}(U^{\dagger})^m {\rm tr} U^m \right\}
\end{equation}
$f({\bf x})$ is the character of the fundamental representation $V_F$.
For $U(3)$ it is
\begin{equation}
  f({\bf x}^m) = x_1^m + x_2^m + x_3^m
\end{equation}
and for $SL(2)$ it is
\begin{equation}
  f({\bf x}^m) = \frac{q^m}{1-q^m}
\end{equation}

Now we perform the group integration for $U(N)$ following
\cite{dutgop} (see also \cite{Dolan:2007rq}).  If we expand out
\begin{equation}
  \exp\left\{ \sum {1 \over m} f({\bf x}^m) {\rm
tr}(U^{\dagger})^m {\rm tr} U^m \right\} \label{start}
\end{equation}
we get
\begin{equation}
  \sum_n \sum_{C_{{\bf i}}\in S_n} \prod_{j=1}^n \left(
  f({\bf x}^j)\right)^{i_j} \frac{1}{j^{i_j}i_j!} \tr(C_{{\bf
      i}} U) \tr(C_{{\bf i}} U^\dagger)
\end{equation}
where $C_{{\bf i}}$ is a partition of $n$ or a conjugacy class of
$S_n$ with $i_1$ 1-cycles, $i_2$ 2-cycles, \dots $i_n$ $n$-cycles.  In
$\frac{1}{j^{i_j}i_j!}$ the $j^{i_j}$ comes from the $\frac{1}{m}$ in
\eqref{start} and the $i_j!$ comes from $\exp(x) = \sum_{k}
\frac{1}{k!}x^k$.

Using the  identity $\tr(C_{{\bf i}} U) = \sum_{R(U(N))} \chi_R(C_{{\bf i}})
\chi_R(U)$ and the group integral
\begin{equation}
  \int [dU] \chi_R(U) \chi_{R'} (U^\dagger) = \delta_{RR'}
\end{equation}
 we get the finite $N$ partition function
\begin{equation}
    \cZ =  \sum_n \sum_{R(U(N))} \sum_{C_{{\bf i}}\in S_n} \prod_{j=1}^n \left(
  f({\bf x}^j)\right)^{i_j} \frac{1}{j^{i_j}i_j!}
  \chi_R(C_{{\bf i}})  \chi_R(C_{{\bf i}})
\end{equation}
Now if we treat ${\bf x}$ as a diagonal matrix (for $U(3)$ we have
$(x_1, x_2, x_3)$ on the diagonal, for $SL(2)$ we have $(q,q^2,q^3,
\dots)$) and use
\begin{equation}
  \prod_{j=1}^n \left(
  f({\bf x}^j)\right)^{i_j} = \tr(C_{{\bf i}}{\bf x}) =  \sum_{\L_1(S_n)} \chi_{\L_1}(C_{{\bf i}})  \chi_{\L_1}({\bf x})
\end{equation}
then we get
\begin{equation}
    \cZ =  \sum_n \sum_{R(U(N))} \sum_{\L_1(S_n)}    \chi_{\L_1}({\bf x})\; C(R,R,\L_1)
\end{equation}
where $C(R,R,\L_1)$ is the number of possible $\tau_{\L_1,R}$
multiplicities in \eqref{eq:finiteNop}, i.e. the number of times
$\L_1$ appears in the symmetric group tensor product $R \otimes
R$.\footnote{$C(R,S,T) = \frac{1}{n!} \sum_{\s \in S_n} \chi_R(\s)
  \chi_S(\s) \chi_T(\s)$ and $\prod_{j=1}^n\frac{1}{j^{i_j}i_j!} =
  \frac{| C_{{\bf i}}|}{n!}$ where $| C_{{\bf i}}|$ is the size of the
  class $C_{{\bf i}}$.}  As representations of $U(N)$, we only sum
over Young diagrams $R$ with at most $N$ rows.

We have treated the global symmetry group here as $GL(\infty)$. 
A further decomposition into irreps. of  $G$ gives 
\begin{equation}
  V_{\L_1}^{GL(\infty)} = \sum_\L V_\L^G \otimes V_{\L,\L_1}
\end{equation}
When we do this we finally see that the operators in
\eqref{eq:finiteNop} provide this counting.
\begin{equation}
    {\cal  Z } =  \sum_n \sum_{R(U(N))}  \sum_{\L(G)}\sum_{\L_1(S_n)} d_{\L,\L_1} \;  \chi_{\L}({\bf x}) \;C(R,R,\L_1)
\end{equation}
where $\chi_{\L}({\bf x})$ is now a $G$ character and $d_{\L,\L_1}$ is
the dimension of $V_{\L,\L_1}$ labelled by the $\tau$ index in
\eqref{eq:finiteNop}.

\section{Worldvolume excitations of  giants  
 and gauge invariant operators}\label{sec:excite}  

\subsection{Worldvolume excitations: review and comments} 

We review and comment on some results from  \cite{djm}
on the worldvolume excitations of half-BPS giant gravitons.   
Consider 3-brane giants expanding in the $AdS^5$. 
Use coordinates $ ( t ,  v_1 , v_2 , v_3 ,v_4 ) $ for the 
AdS where we have a metric 
\bea 
ds^2 = - ( 1 + \sum_{k=1}^{4} v_k^2    ) dt^2 + 
L^2  ( \delta_{ij}  + { v_i v_j  \over  ( 1 + \sum_{k} v_k^2  )  }) dv_i dv_j 
\eea 
$L$ is the $AdS_5$ or $S^5$-radius. 
The $S^5$ can be described in analogous coordinates 
\bea 
ds^2 = L^2 [ ( 1- \sum_{k=1}^4  y_k^2 ) d \phi^2 + ( \delta_{ij} +
  {  y_iy_j \over 1 - \sum_k y_k^2 } ) dy_i dy_j  ] 
\eea 
In global coordinates the AdS metric is 
\bea 
ds^2 = - ( 1 + {r^2 \over L^2 } ) dt^2 +
 { dr^2 \over ( 1 + { r^2 \over L^2} ) }   + r^2 d\Omega_3^2 
\eea 
It is also useful to write the $S^5$ metric as  
\bea 
ds^2 = L^2 ( d \theta^2 + \cos^2  \theta d \phi^2 + \sin^2 \theta d\Omega_3^2 ) 
\eea  
The AdS-giant graviton solution has 
\bea 
&&\phi = \omega_0 t \cr 
&& \omega_0 = { 1 \over L } \cr 
&& P_{\phi } = N ( { r_0 \over L  } )^2  
\eea 
and the half-BPS property guarantees the energy is 
$ E = { P_{\phi } \over  L } $. 
The brane worldvolume coordinates are $ \tau , \sigma_1 , \sigma_2 , 
\sigma_3 $. The coordinate $ \tau $ is identified with the global time 
$t$. The $ \sigma_1, \s_2 , \s_3 $ are identified with 
angles in $AdS$. 

 The  fluctuations are expanded as  
\bea 
&&  r  = r_0  + \epsilon ~ \delta r ( \tau , \sigma_1 , \sigma_2  , \s_3 ) \cr 
&& \phi = \omega_0 \tau + \epsilon  ~ \delta \phi ( \tau , \s_1 , \s_2 , \s_3 ) \cr&& y_k = \epsilon ~ \delta {y_k} ( \tau , \sigma_1 , \sigma_2 , \sigma_3 ) 
\eea

These perturbations are expanded in spherical harmonics. 
\bea\label{sphharm}  
&& \delta r ( \tau , \sigma_i ) = \tilde \delta r ~  e^{-i \omega \tau } Y_l ( \tau , \sigma_i ) \cr 
&& \delta \phi ( \tau , \sigma_i ) = \tilde \delta  \phi ~  e^{ -i \omega \tau } 
Y_l   ( \tau , \s_i ) \cr 
&& \delta y_k = \tilde \delta y_k ~  e^{ -i \omega \tau }  Y_l ( \tau , \s_i ) 
\eea 
The $( \phi , y_k ) $ are coordinates for the sphere $S^5$. 
The $Y_l $ are spherical harmonics on $S^3 \subset AdS_5 $.
 They are symmetric traceless 
representations of $SO(4)$. They have a quadratic Casimir 
$ { l ( l +2 )  } $ for the symmetric traceless representation 
of dimension $ (l+1)^2 $. 
The frequencies of these oscillations are  calculated from 
the linearized equations of motion of the brane actions 
\bea 
S = S_{ DBI } + S_{ CS } 
\eea 
They lead (after a small simplification of expressions in \cite{djm}) 
 to three solutions 
\bea\label{eigenfreq}  
&& \omega_- = { l \over L } \cr 
&& \omega_+ = { l + 2 \over L } \cr 
&& \omega =  { l +1 \over L }  
\eea 
The modes with frequencies  $ \omega_{\pm}$
 are related to linear combinations $ \tilde \delta r , \tilde \delta \phi $. 
The frequency  $ \omega $ is related to four modes $ \tilde \delta v_m$ 
which transform in the fundamental of $SO(4)$ in $SO(6)$.  
It is very interesting that these are all integer 
multiples of the AdS-scale and approach $ \omega = l/L  $
in the large $l$ limit. Note also 
that $ \omega $ is the frequency for oscillations 
in $ t $ , the global time of AdS. The energies of 
the fluctuating giant gravitons are given by 
$ E = { n \over L  } + \omega $ where $n$ is the angular momentum 
of the background giant. The energy is related to 
scaling dimension in the dual CFT \cite{wit}.  
 These energy spacings 
in integer units of ${ 1\over L } $ are precisely the sort 
of spacings we get in free Yang Mills theory. 
Taking large angular momentum limits as
 a way to reach a classical regime where 
strong and weak coupling coupling can be compared directly 
is familiar from \cite{gkpi}.

The $Y_{l,m}$ are representations of $SO(4)$. 
Specifying the eigenvalues of the Cartan amounts to 
fixing two spins $S_1 , S_2 $. The $SL(2)$ sector 
of gauge theory operators we considered, involving 
multitraces of $ \d_{1+i2 }^S X^n   $ corresponds to 
rotations in a fixed plane.  This means that 
in each space of spherical harmonics of given $l$
we are looking  at a single state. 
Now if we consider  a second quantization in the field theory 
of the branes, we would introduce a Fock space generator  
$ \alpha_{l}^{\dagger}   $ for each spherical  harmonic. This has energy $l/L $
above the background energy of the brane. 
General states look like 
\bea\label{giantexcits}
\alpha_{ 1 }^{\dagger ~ k_1} \alpha_{2}^{\dagger ~  k_2} \cdots | 0 \ra 
\eea 
The number of states at excitation energy $k$ is the 
number of ways of writing $k = k_1 + 2k_2 + \cdots  = \sum_{i} k_i l_i $
 which is the number of partitions of $k$.
When we restore the full $SO(4)$ we have states of the form  
\bea\label{giantexcits2}
\alpha_{ l_1 , m_1  }^{\dagger}  \alpha_{l_2 , m_2 }^{ \dagger } 
\cdots | 0 \ra 
\eea 
In this case it is useful  to restrict attention to the 
symmetric traceless representations $[k]$  of $ SO(4) $ with 
excitation energy equal to $k$. In this case, the 
number of excited states of total energy  $L_0 = n + k$ is again 
given by partitions of $k$. 
In the discussion below we will show that that there is an easy 
way to get these states from the gauge theory. In greater generality 
we should consider states of the form 
\bea\label{giantexcits3}
\alpha_{ l_1 , m_1 , I_1 }^{\dagger}  \alpha_{l_2 , m_2, I_2  }^{ \dagger } 
\cdots | 0 \ra 
\eea 
where $I$'s are indices running from $1$ to $6$ which label the 
six eigenmodes built from (\ref{sphharm}). 
 Four of these are in the fundamental of 
$SO(4) \subset SO(6)$.  
The fact that the excitation energies are  
spaced in units of ${ 1 \over L } $ (rather than in units of 
the brane size) was a bit of 
a surprise, discussed at length in \cite{djm}. 
An important point  is that 
the kind of integer spacing in (\ref{eigenfreq})
 is exactly what we have in free Yang Mills
limit of the dual CFT.  We will see below that this Fock space
structure of orthogonal states  emerges indeed 
from the construction of gauge invariant operators in the 
free dual Yang Mills theory. A connection between excited 
giant gravitons and the formulae for excitation energies (\ref{eigenfreq}) 
was made in \cite{BHLN}. The unravelling of the Fock space structure
of giant graviton worldvolume field theory  from gauge invariant operator
 counting given below is new.

\subsection{Comparison to gauge invariant operators} 

We have constructed,  in section \ref{sec:SL2},  
the lowest weights of the $SL(2)$ sector 
by mapping states 
\bea 
A_{ a_1}^{ \dagger}  A_{a_2}^{\dagger}  \cdots A_{a_l }^{\dagger}  | 0 \ra 
\eea 
in an oscillator construction of $SL(2)$ to gauge theory operators. 
The index  $a$ transforms in the hook representation $ [n-1,1]$ of $S_n$. 
The $A^{\dagger} $'s are
 bosons so we are looking at the symmetric tensor product 
 of the hook. 
These were constructed as lowest weight 
states generated  by $ P_{1 \dot 1 } $ 
which forms part of the  $SO(4,2) $ conformal algebra. These excitations 
correspond to exciting one spin inside AdS (for more details 
on the geometry of multiple spins see for example \cite{steftsey}
in the context of spinning strings), hence to states of the form 
(\ref{giantexcits}).   After describing how to lift this to 
more general $SO(4,2)$ states, we will show that the counting
in the case of single giants agrees with the bulk analysis reviewed above. 
Note for now that the above states transform in 
 $ Sym ( V_H^{\otimes k} ) $ of $S_n$. 

When we consider the full  
$SO(4,2)$ symmetry, we have additional generators $K_{\lambda } $ 
forming the fundamental of $SO(4)$. Correspondingly we have $P_{\lambda}$ 
transforming in the fundamental of $SO(4)$.  When we consider lowest weight 
states annihilated by all the $K_{\lambda } $, we have states of the form  
\bea\label{oscso4}  
A^{\dagger}_{ a_1 \l_1 }  A^{\dagger}_{a_2 \l_2} \cdots 
A^{\dagger}_{a_k \l_k  } | 0 \ra 
\eea 
Among these LWS are those transforming in the symmetric traceless 
representation of $SO(4)$ associated with the symmetric Young diagram $[k]$
and with energy $ L_0 = n + k $. 
As discussed in  \ref{sec:SO42}, 
these are a simple class of states which do not require 
projecting out of states due to the equations of motion, 
which require setting $ P_{\lambda }P_{\lambda } $ to zero.  
Since the $\lambda$'s are symmetrised, and the $A^{\dagger}$ are 
bosons, the indices $a_1, a_2 , \cdots , a_l$ are symmetric, 
i.e. we have the symmetric $k$-fold tensor power of the hook representation 
$[n-1,1]$ of $S_n$. Orthonormal states in this sector are then written as 
\bea\label{so4excit}  
C_{\l_1 \cdots \l_k }^{[k], M_{[k]} }
C_{a_1 ... a_k }^{ \L_1 ,[k] ,  m_{\l_1} , \tau' }  
    A^{\dagger}_{ a_1 \l_1 }  A^{\dagger}_{a_2 \l_2} \cdots 
A^{\dagger}_{a_k \l_k  } | 0 \ra 
\eea
The first Clebsch's are for the symmetric traceless of $SO(4)$ which are 
precisely the representations we discussed under \eqref{giantexcits2}.
 The second Clebsch have been discussed 
before in Section 2. They decompose the $ Sym ( V_{H}^{\otimes k } )$ 
into irreps. of  $ \L_1$ of  $S_n$. 
 When we form gauge invariant operators as 
in Section \ref{sec:gaugeN} there are constraints relating 
$\L_1 $ to the $U(N)$ Young diagram $R$ which organises the traces. 
This representation $R$ in the half-BPS case 
allows a map to the type of giant \cite{cjr}. Young diagrams 
with a few  (order $1$) long (order $N$ for example) 
 columns map to sphere giants. Those with a few long rows map to AdS giants.
Non-abelian worldvolume symmetries emerge when we have rows or columns 
of equal length.  This map also works for open string 
excitations and there are elegant tests involving the counting of
 states which are sensitive to the presence of non-abelian symmetries
\cite{bbfh,rob}.

Consider  Young diagrams of the form  
$R = [n ]$ which correspond to single AdS giants of 
angular momentum $n$. 
Recall that the gauge invariant operators are labelled 
by $ R , \L_{n+ k  } , M , \L_{1} , \tau , \tau_{\L_1,R} $.  
$R$ is a $U(N)$ irrep. $\L_{ n+k } $ is the lowest weight 
of the $SL(2)$ which is completely determined by the 
excitation energy  $l$. $M$ labels states in $\L_{ n+ k  } $. 
 $ \L_1 $ is an irrep. of $S_n$. 
$\tau$ runs over the multiplicity of $ \L_1 $ in the  
symmetric tensor product of the hook representation. 
$ \tau_{\L_1,R} $ runs from $1$ to $ C ( R , R , \L_1 ) $. 
For fixed $R$  the multiplicity of LWS  
is 
\bea\label{smfxdr}  
\sum_{ \L_1} C ( R , R , \L_1 ) Mult ( \Sym ( V_H^{\otimes k} ) , \L_1 ) 
\eea    
By summing over states for fixed $R$ we can get excited states 
of a fixed type of giant worldvolume.  
In particular we are interested in $ R = [n ]$. 
The inner tensor product of $ R $ with itself only contains 
the identity rep. $ \L_1 = [n] $. So the number of lowest weights 
at level $k$ is just the multiplicity of 
$[n]$ in the symmetric tensor product of the hook. 
We have a generating function for this derived in Section 3.  
The generating function including the descendants, 
 is (using  (\ref{symsum}) or (\ref{conj}) ) 
\bea\label{fckcount}  
{ 1 \over ( 1 - q )  ( 1 -q^2 ) ( 1 - q^3 ) \cdots ( 1 - q^n ) } 
\eea 
The coefficient of $q^k$ is the number of partitions of 
$k$ with no part bigger than $n$. Note that $n$ is the 
number of boxes in the Young diagram describing the giant. 
For the semiclassical approximation  of giant brane worldvolume 
 to be valid, this is of order $N^{\alpha} $ (for $ \alpha $ close to $1$), 
 $k$ is the excitation on the brane worldvolume, which we are treating 
in a linearized approximation, so we certainly want that to be small 
compared to $n$. When $k$ is smaller than $n$, the above just counts 
unrestricted partitions of $k$. This matches the counting of Fock 
space states in (\ref{so4excit}). 

Hence, in the regime of interest, where $k$ is much bigger 
than one (so we can expect GKP \cite{gkpi} type arguments to be valid)   
but smaller than the energy of the brane, the above counting 
of partitions of $k$ is exactly what we are getting from quantizing 
a class of vibrations of the AdS giant.  
Using this emergence of Fock space structures from 
the counting of states in the tensor product of $ \Sym( V_H^{\otimes k } ) $ 
we therefore find the correct counting of 
gauge theory operators which correspond to 
states of the form \eqref{giantexcits} and \eqref{giantexcits2}
with energy $ L_0 = n + k $ and  
 with a single spin $k$ in the case \eqref{giantexcits}
or with  $SO(4)$ representation $[k] $ for \eqref{giantexcits2}.

In fact we can also see where the six different species of 
oscillations could come from. In the above discussion we 
have been considering BPS giants built from Schur 
polynomials of $ X  = X_1 + i X_2$ and then perturbed
by replacing $X$ with   derivatives $ P_{\lambda } $ acting on  $X$, 
 of the form $ P_{\lambda }^* X$.  
We could also 
consider powers of $ P_{\lambda } $ acting on $ X_i $ 
(with $i=3,4,5,6$ ) replacing the $X$. And finally we can consider 
powers of $P_{\lambda }$ acting on     $X^{\dagger}$ as the impurities.
So in all we have six types of impurities $P_{\lambda}^* X , P_{\lambda}^* X^{\dagger} , P_{\lambda}^* X_i $. 
These correspond to six  sets of gauge invariant 
operators  matching states with the right energies 
of the form (\ref{giantexcits}), which 
come, in the spacetime worldvolume analysis to exciting quanta of  
$ \delta r ,  \delta v_m , \delta \phi $  excitations.
Given the simplicity of $\omega_- $ we would expect that they correspond to 
the simplest construction in gauge theory, namely using 
$P_{\lambda}^* X $ impurities, which they match precisely in energy. 
If we consider the states in \eqref{giantexcits3}  and restrict 
to the case where all the impurities are of the same type and the 
$SO(4)$ representation is $[k]$ with the energy being $ E = n +k $, 
then the above discussion extends easily to give the corresponding gauge 
theory duals.  A complete account of the case with mixed impurities 
will be left for the future.

\subsection{Comments}

There are many interesting extensions of the above discussion 
which could be considered. We have chosen the simplest 
$R$ of the form $[n]$ which correspond to AdS giants. 
If we consider $R = [n_1 , n_2 ] $ and sum over $\L_1$ as in 
(\ref{smfxdr})  this should correspond to excitations in spacetime of 
multiple-giants described by a $U(2)$ (if $n_1= n_2 $ ) 
or $U(1) \times U(1) $ (if $n_1 \ne n_2 $ ) worldvolume DBI gauge theory. 
A similar simple counting of states  holds true  for excitations of S-giants
\cite{djm}.  They will be associated to spherical harmonics of an $SO(4) $ 
in the $SO(6)$. So we expect that excitations in the gauge theory from the 
$SO(6)$ sector should also have this kind of free field 
counting in an appropriate large angular momentum  limit.
The $SO(4) \subset SO(4,2)$ excitations considered in 
(\ref{so4excit}) also exist for $R = [1^n]$. They should correspond 
to excitations of sphere giants, but it is not obvious to us  
how a Fock space structure emerges from considering their 
motions in the transverse AdS. It will be interesting to 
clarify this puzzle. 

Note that we are making here a comparison between 
zero coupling in Yang Mills to spacetime calculations 
dual to strong coupling Yang Mills. This  works best 
for large angular momenta where $ l $ is large so that the frequencies 
can all be approximated by $ \omega = l $, but 
smaller than $n$ which is the large angular momentum of the giant.    
This gives a different context of excitations of giant gravitons, 
where the basic idea of large quantum numbers allowing 
strong to weak coupling comparisons  \cite{gkpi} continues to apply.  
Here the parameters $N , k , n $ are all large. 

There have been earlier discussions of supersymmetric states
obtained from the quantization of moduli spaces of giants
and the comparison with gauge theory counting \cite{bglm,ms,calsil,ggkm}.
In the discussion above we have been interested in all 
the excitations in the free theory of a given half-BPS giant. 
A subset of these will be supersymmetric but a lot of the states 
will be non-supersymmetric. We expect that, in analogy with 
discussions of semiclassical strings \cite{gkpi,steftsey} 
appropriate limits of large quantum numbers can be used to 
compare non-supersymmetric states. 
The new technical  ingredient in the above treatment is 
the use of a diagonal basis of gauge theory operators 
at finite $N$, where the label $R$ allows the identification 
of the giant in question, and additional global symmetry labels 
help the map to objects in spacetime. The use of symmetric group 
data in organising the multiplicities of states for fixed $R$ and 
fixed global symmetry quantum numbers shows  the 
emergence, in the limit of large $n$,   of Fock space counting 
from properties of symmetric group decompositions 
such as $ \Sym( V_{H}^{ \otimes n }) $. At finite $n$ we have a cut-off 
Fock space.

\section{One-loop mixing in $\cN=4$}\label{sec:1-loop} 

In \cite{Brown:2008rs} the one-loop mixing of the Clebsch-Gordan basis
introduced in \cite{bhr} for the $G = U(2)$ sector of $\cN=4$ super
Yang-Mills was analysed.  These operators only mix if the $U(N)$
representations specifying their multi-trace structures are related by
the repositioning of a single box of the Young diagram.  Here we find
the same result for the full $PSU(2,2|4)$ sector, using our general
characterisation of multi-trace operators with arbitrary global
symmetry.

The complete one-loop non-planar dilatation operator is given by
\cite{beisert}
\begin{equation}
  D(g) = D_0 - \frac{g_{YM}^2}{8\pi^2} H + \cO(g_{YM}^3)
\end{equation}
where
\begin{equation}
  H = \sum_{j=0}^\infty h(j) (P_j)^{AB}_{CD} :\tr ([W_A ,\tilde W^C][W_B ,\tilde W^D]): \label{fulloneloop}
\end{equation}
$(\tilde W^C)^i_j$ is the derivative $\frac{d}{d (W_C)^j_i}$.  $h(j)
\equiv \sum_{k=1}^j \frac{1}{k}$ are the harmonic numbers and $P_j$ is
the projector for $V_F \otimes V_F = \oplus_j V_j$.  For $SL(2)$ and
$PSU(2,2|4)$ $V_j$ appears with unit multiplicity in $V_F^{\otimes 2}$
(cf. \eqref{mkn} where $m(j,2) = 1$)\footnote{In the $SL(2)\times S_2$
  decomposition of $V_F^{\otimes 2}$, the symmetric representation
  $V_{[2]}^{S_2}$ appears with even $j$ and the
  antisymmetric $V_{[1,1]}^{S_2}$ with odd $j$.}. The dilatation operator
separates out $V_F^{\otimes 2}$ in $V_F^{\otimes n}$ and then projects
onto it with the factors in \eqref{fulloneloop}.

The action of the dilatation operator has been analysed in the planar
limit for single traces using the Bethe Ansatz (see for example
\cite{Minahan:2002ve}\cite{Beisert:2003yb}).  In the non-planar limit
multi-trace operators can join and split \cite{Bellucci:2004ru}.  We
will find that the mixing is neatly constrained if we organise the
multi-trace operators using $U(N)$ representations as we have in
\eqref{eq:finiteNop}.

The action of $H$ on $\tr(\a W_{m_1} \cdots W_{m_n})$ is compactly
written by introducing an extra index, tracing in $V_N^{n+1}$ rather
than $V_N^n$.  The extra index encodes awkward contractions in the
action of the dilatation operator.
\begin{align}
&    :\tr ([W_A ,\tilde W^C][W_B ,\tilde W^D]): \tr(\a W_{m_1} \cdots W_{m_n}) = \nn \\
&  \frac{1}{(n-2)!}\sum_{\s \in S_n} \delta^C_{m_{\s(n-1)}} \delta^D_{m_{\s(n)}}\sum_{\rho_1,\rho_2 \in S_{n+1}} f(\rho_1,\rho_2) \tr_{n+1} \left( \rho_1\s^{-1}\a\s \rho_2 W_{m_{\s(1)}} \cdots W_{m_{\s(n-2)}}W_A W_B \bI_N \right) \nn
\end{align}
$\bI_N$ is the $N \times N$ identity matrix.  $f(\rho_1, \rho_2)$ is
only non-zero on the $S_3$ subgroup of $S_{n+1}$ that permutes the
$n-1$ and $n$ indices, where the derivatives act, and the new $n+1$
index.  Its non-zero values give the four terms of the commutators in
\eqref{fulloneloop}.
\begin{align}
  f(\;(n-1,n)\;,\; (n,n+1) \;) &= 1 \nn \\
  f(\;(n-1,n+1)\;,\;(n,n+1) \;) &= -1 \nn \\
  f(\;(n,n+1)\;,\;(n-1,n+1) \;) &= -1 \nn \\
  f(\; (n,n+1) \;,\; (n-1,n) \;) &= 1 
\end{align}

If we introduce the projector we find
\begin{align}
&  \sum_{j=0}^\infty h(j) (P_j)^{AB}_{CD}  :\tr ([W_A ,\tilde W^C][W_B ,\tilde W^D]): \tr(\a W_{m_1} \cdots W_{m_n})  = \sum_{\rho_1,\rho_2 \in S_{n+1}} f(\rho_1,\rho_2)\nn \\
&\frac{1}{(n-2)!}\sum_{\s \in S_n}\sum_{j=0}^\infty h(j) \tr_{n+1}\left( \rho_1\s^{-1}\a\s \rho_2 W_{m_{\s(1)}} \cdots W_{m_{\s(n-2)}} P_j \left(  W_{m_{\s(n-1)}} W_{m_{\s(n)}} \right) \bI_N \right) \nn
\end{align}  
Now consider the action on our gauge-invariant operator \eqref{eq:finiteNop}
\begin{align}
&H   \cO^{\L,M_\L, \L_1, \tau,R,\tau_{\L_1,R}}  = \frac{1}{(n-2)!} \sum_{\rho_1,\rho_2 \in S_{n+1}} f(\rho_1,\rho_2) \; C^{ \tau_{\L_1,R}}\;{}^{ \L_1}_{ m_{\L_1}}\;{}^R_i\;{}^R_j\;\;  \sum_{\a\in S_n} D^R_{ij} (\a)\nn \\
&\sum_{j=0}^\infty h(j) C^{\vec{m}}_{\L,M_\L,\L_1,m_{\L_1},\tau} \tr_{n+1}\left( \rho_1\a \rho_2 W_{m_{1}} \cdots W_{m_{n-2}} P_j \left(  W_{m_{n-1}} W_{m_{n}} \right) \bI_N \right) \  \label{dilonop}
\end{align}  
Here, using properties of our operators, all the $\s$ actions cancel.

To encapsulate the action of the projector we rewrite the covariant
decomposition of $V_F^{\otimes n}$ in terms of $V_F^{\otimes n-2} \otimes
V_F^{\otimes 2}$. We unclutter the notation by defining $\ket{ {\bf \L}}
\equiv \ket{\L,M_\L,\l,m_{\l},\tau}$ for the covariant basis.
\begin{align}
  \ket{{\bf \L}} =& \sum_{\vec{m}}  C^{\vec{m}}_{{\bf \L}} \;\; \sum_{{\bf \L}^{n-2},{\bf \L}^{2}} \;\; C_{\vec{m}^{n-2}}^{{\bf \L}^{n-2}}\;\;C_{\vec{m}^{2}}^{{\bf \L}^{2}} \;\;  \ket{{\bf \L}^{n-2}} \otimes   \ket{{\bf \L}^{2}} \nn \\
 =& \sum_{{\bf \L}^{n-2},{\bf \L}^{2}} \;\; \braket{{\bf \L}^{n-2},{\bf \L}^{2}}{{\bf \L}} \;\;  \ket{{\bf \L}^{n-2},{\bf \L}^{2}} 
\end{align}
$\ket{{\bf \L}^{n-2}}$ lives in $V_F^{\otimes n-2}$ while $\ket{{\bf
    \L}^{2}}$ lives in $V_F^{\otimes 2}$.  $\vec{m}^{n-2} = (m_1, \dots,
m_{n-2})$ and $\vec{m}^{2} = (m_{n-1}, m_{n})$.

The projector $P_j$ in \eqref{dilonop} projects onto ${\bf \L}^{2} =
j$.  The one-loop two-point function is then

\begin{align}
&   \corr{  (\cO^\dagger)^{\L',M'_{\L'}, \L'_1, \tau',R',\tau'_{\L_1',R'}}\;\; H \;\; \cO^{\L,M_\L, \L_1, \tau,R,\tau_{\L_1,R}} } \nn \\
& = \frac{1}{(n-2)!} \sum_{\rho_1,\rho_2 \in S_{n+1}} f(\rho_1,\rho_2) \; C^{ \tau_{\L_1,R}}\;{}^{ \L_1}_{ m_{\L_1}}\;{}^R_i\;{}^R_j\;\;C^{ \tau'_{\L_1',R'}}\;{}^{ \L_1'}_{ m'_{\L_1'}}\;{}^{R'}_k\;{}^{R'}_l\;\;  \sum_{\a,\a'\in S_n} D^R_{ij} (\a)\; D^{R'}_{kl} (\a') \nn \\
&  \sum_{{\bf \L}^{n-2},{\bf \L}^{2} = j}   h(j) \; \braket{{\bf \L}'}{{\bf \L}^{n-2},{\bf \L}^{2}} \;\braket{{\bf \L}^{n-2},{\bf \L}^{2}}{{\bf \L}} \; \tr_{n+1}\left( \rho_1\a \rho_2 \a'\, \bI_N^{n+1} \right)\label{finaloneloop}
\end{align}
The trace can be expressed as a sum over $(n+1)$-box representations
$T$ of $S_{n+1}$ and $U(N)$ with at most $N$ rows.
\begin{equation}
   \tr_{n+1}\left( \rho_1\a \rho_2 \a'\, \bI_N^{n+1} \right) =
\sum_{T \vdash n+1} \chi_T( \rho_1\a \rho_2 \a' ) \Dim T
\end{equation}
The $\a$ and $\a'$ sums in \eqref{finaloneloop} force $T$ to reduce to
both $R$ and $R'$ for the $S_n$ subgroup of $S_{n+1}$.  Since $T$
reduces on its $S_n$ subgroup to those Young diagrams with a single
box removed from $T$, $R$ and $R'$ must be related by the
repositioning of a single box for this one-loop two-point function not
to vanish.  This analysis is pursued in more detail in
\cite{Brown:2008rs}.

The one-loop non-planar mixing of this complete basis of multi-trace
operators is therefore highly constrained.  Although the operators are
not diagonal at one-loop, their very limited mixing suggests they are
close to the eigenstates.  It would be particularly interesting to
find the sixteenth-BPS operators at one loop and gain an understanding
of the counting of black hole entropy, along the lines of
\cite{Janik:2007pm,ggkm}.

\section{Summary and Outlook} 

We have given a general construction of diagonal gauge invariant
operators for a $U(N)$ gauge theory with global symmetry $G$ in the
free field limit.  The Clebsch-Gordan coefficients for the $ G \times
S_n$ decomposition of $V^{\otimes n }_F $, where $V_F$ is a fundamental
representation of $ G$, play a crucial role. We have exploited the
result of \cite{bhr} which showed that once the covariant correlators
are brought to a standard form, which we have called the ``canonical
covariant form'' then the gauge invariant diagonalisation follows
using Clebsch-Gordan coefficients of $S_n$. The $G \times S_n$
decomposition contains representation labels $ \L , \L_1$ of $ G $ and
$S_n$ respectively.  We showed in this paper how to use the
corresponding Clebsch-Gordan coefficients, obeying standard Clebsch
orthogonality properties (\ref{orthogLMI}), to construct operators
with correlators of the canonical covariant form (\ref{finalorh}).
 
The construction of gauge invariant objects uses a representation label 
$R$ corresponding to $U(N)$ and its Schur-Weyl dual $S_n$. This 
label appears in the simplest set-up in the half-BPS sector \cite{cjr}, 
and is interpreted in terms of giant gravitons. The final step 
of going from canonical covariant form to gauge invariant diagonal form 
uses the Clebsch-Gordan coefficients for the $S_n$ inner tensor
 product $ R \otimes R \rightarrow \L_1 $. We showed that  the 
construction of a diagonal basis of gauge invariant operators 
matches the counting of gauge invariant operators done using 
Matrix Model techniques \cite{dutgop}. 
 
As special cases we have considered $ G = U(M) , SL(2) , SO(6)$ which
are relevant to specific sectors of $ \cN =4 $ SYM theory. In the case
of $U(M)$ we have shown that the multi-matrix diagonalisation result
of \cite{bhr} contains a formula for the Clebsch-problem of $
U(M) $ decomposition of $V_M^{\otimes n } $ in terms of the symmetric
group data of branching coefficients.  For $SL(2)$ we have shown how
the $SL(2) \times S_n $ Clebsch-problem for the $n$-fold tensor
product of the discrete series representation spanned by $ X , \d X ,
\d^2 X ... $ can be solved by considering one energy level at a time,
labelled by $k$, the total number of derivatives involved in the
$n$-fold tensor product.  The total number of lowest weight states
appearing at fixed $k$ can be neatly described in terms of oscillator
constructions of $SL(2)$.  This leads to a mapping of the problem of
diagonalising the multiplicity of LWS at fixed $k$ into a problem
involving the $S_n \times S_k$ decomposition of the $k$-fold tensor
power of a hook representation of $S_n$ of dimension $n-1$.  This $S_n
\times S_k$ problem has some surprising Fock space structures in the
large $n, k $ limit. These structures have been used to identify gauge
invariant operators with energies and multiplicities matching those
appearing in earlier work on the  excitations of giant
gravitons computed from the point of view of a worldvolume analysis
\cite{djm}.

We expect that further  investigations on the diagonalisation 
of gauge invariant operators will allow more detailed  comparisons 
between excitations of giant gravitons in spacetime and gauge theory
operators. Comparisons going beyond free fields in gauge theory 
and beyond the leading semiclassical approximations 
in giant gravitons will also be instructive. The formula for 
the 1-loop dilatation operator acting on our basis, given in 
Section (\ref{sec:1-loop}), is a step in this direction.

\vskip.5in 

{ \bf Acknowledgements } We thank  Sumit Das, 
 Robert de Mello Koch, Yusuke Kimura, David Turton  for discussions. SR is
 supported by an STFC Advanced Fellowship. PJH is supported
 by an EPSRC Standard Research Grant EP/C544250/1.  TWB is on an STFC
 studentship.

\vskip.8in

\begin{appendix}

\section{Clebsch-Gordan decomposition  for $V_{H}^{\otimes k  } $}\label{sec:symmCG}
We will collect several useful facts about the decomposition 
into $S_n \times S_k$ representations of the $k$-fold 
tensor product $V_H^{\otimes k } $ of the hook representation 
$V_H$ of $S_n$ associated with the Young diagram $[n-1,1]$. 

\subsection{Multiplicities from characters }\label{app:hookdecomp} 

The $k$-fold tensor product decomposes as follows
\bea 
V_H^{ \otimes k } = \bigoplus_{\L_1,\L_2} V_{ \L_1 } \otimes V_{ \L_2 } \otimes 
  V_{ \L_1 , \L_2 }
\eea 
Here $  V_{ \L_1 } $ is an irrep of $S_n $, $ V_{ \L_2 } $ is an 
irrep.  of  $ S_k$  , and $ V_{ \L_1 , \L_2} $ is an irrep of 
$ \Com ( S_n \times S_k  ) $, the algebra commuting with $ S_n \times S_k $ 
in the  $V_H^{ \otimes k }$.

The dimensions $ d_{ \L_1, \L_2} $ of $ V_{ \L_1 , \L_2 }$ 
appear in the oscillator construction of LWS (lowest weight states) 
in the tensor products of the fundamental $SL(2)$ representation. 
We can calculate these dimensions using characters of $S_n $ and $S_k$. 
\bea 
d_{ \L_1, \L_2} = tr_{ V_H^{\otimes k } } \left( { P_{\L_1}\over d_{ \L_1}} 
 \otimes { P_{ \L_2} \over d_{ \L_2} }  \right) 
\eea 
The projectors $P_{ \L_1} , P_{ \L_2} $ are given by 
\bea 
{ P_{ \L_1} \over d_{ \L_1} }  = 
{ 1 \over n! }  \sum_{ \sigma \in S_n }     \chi_{ \L_1} ( \s )  \s
\eea 
and 
\bea 
{ P_{ \L_2} \over d_{ \L_2} }  = 
{ 1 \over k! }  \sum_{ \tau \in S_k }     \chi_{ \L_2} ( \tau  )  \tau  
\eea 
Hence we can write 
\bea 
&& d_{ \L_1, \L_2} 
= { 1\over k! } \sum_{ \tau \in S_k   } \chi_{ \L_2} ( \tau ) 
 { 1 \over n! } \sum_{ \sigma  \in S_n   } \chi_{ \L_1} ( \sigma  ) 
  tr_{V_H^{\otimes k} } (  \tau   \otimes \sigma  ) \cr  
&& = { 1\over k! } \sum_{ \tau \in S_k   } \chi_{ \L_2} ( \tau )
 { 1 \over n! } \sum_{ \sigma  \in S_n   } \chi_{ \L_1} ( \sigma  )
    \prod_i   ( tr_{V_H } ( \sigma^i   ))^{ c_i ( \tau ) }  
\eea 
For computer code to calculate this multiplicity see Section \ref{sec:code}.
Here $c_i ( \tau )  $ is the number of  cycles of length $ i  $
in the permutation $ \tau $. 
To see this note that 
\bea 
&& \bra{ a_1 .. a_k } \sigma \ket{   a_1 .. a_k } = \braket{ a_1 .. a_k }{\sigma(a_1) \cdots 
\sigma ( a_k ) } \cr 
&& \bra{ a_1 .. a_k } \tau  \ket{   a_1 .. a_k }= \braket{ a_1 .. a_k }{ a_{ \tau (1) } 
 \cdots 
 a_{ \tau (k) }   } \cr 
&& \bra{ a_1 } \sigma\ket{ a_2} = D^{H}_{ a_1 a_2 } (  \s ) 
\eea

\subsection{Clebsch-Gordan coefficients}

For the Clebsch-decomposition of $V_H^{ \otimes k } $ into $ \L_1 (
S_n ) \otimes \L_2 ( S_k ) $, the first thing we need is the
multiplicities. We also need in Section 2 the properties of the
Clebsch-Gordan coefficients for the case $ \L_2 = [k] $.  Here we
state some general properties valid for any $ \L_2 $.

For the basic Clebsch problem of coupling a pair of irreps to a third
$ R \otimes S \rightarrow T $ we have previously used formulae of the
type $ DDD = C C $ and $ DD C = D C $ derived for example in
\cite{hamermesh}\footnote{ A minor notational point is that we are
  using the symbol $C$ for Clebsch in this paper rather than $S$ as in
  \cite{bhr} and \cite{hamermesh}.}.  These were derived by inserting
complete sets of states etc.  Similar techniques lead to similar
equations, which allow us , given the matrix elements of $S_n$
irreps., to compute the Clebsch.

The following is the analog of $DDC  = DC  $
\bea 
&& \sum_{ a_1 , \cdots , a_k } D^{H}_{b_1 a_{\tau (1)} } ( \s ) 
D^{H}_{b_2 a_{\tau (2)} } ( \s )  \cdots D^{H}_{b_k a_{\tau (k)} } ( \s ) 
C_{a_1 ... a_k }^{ \L_1 , \L_2 , m_{ \L_1 } , m_{ \L_2} ; ~ \tau_{ \L_1 , \L_2 } }
\cr 
&& = C^{\L_1 , \L_2, m'_{\L_1 }
 , m_{ \L_2}' ; ~  \tau_{\L_1 , \L_2 } }_{b_1 \cdots b_k }    
   D^{ \L_1}_{ m'_{\L_1} m_{ \L_1} } ( \s )    
D^{ \L_2}_{ m'_{\L_2} m_{ \L_2} } ( \tau ) 
\eea 

The following is the analog of $ DDD = CC  $. 
\bea 
&& \sum_{ \sigma \in S_n } \sum_{ \tau \in S_k } 
D^{\L_1 }_{ m_{ \L_1 } , m'_{\L_1} } ( \s )  
D^{\L_2 }_{ m_{ \L_2 } , m'_{\L_2} } ( \tau  )
~~~ D^{H}_{ a_1 b_{ \tau (1) } }( \s )  \cdots    
D^{H}_{ a_k b_{ \tau (k) } } ( \s )   \cr 
&& = \sum_{ \tau_{\L_1 , \L_2 }  } 
     C^{ \tau_{\L_1 , \L_2} , \L_1 , \L_2 , m_{ \L_1 } ,
 m_{ \L_2 } }_{ b_1 ... b_k } 
   C^{ \tau_{\L_1 , \L_2} , \L_1 , \L_2 , m_{ \L_1 }' ,
 m_{ \L_2 }' }_{ a_1 ... a_k }  
\eea

\subsection{Symmetrised Clebsch from ordinary Clebsch } 

Now we specialise to give the properties for $ \L_2 = [k] $
which involve the symmetrised Clebsch.  
The symmetrised Clebsch-Gordan coefficients
  with the properties used 
in Section 2  can be obtained from the Clebsch-Gordans
 for ordinary tensor products $V_H^{ \otimes k }  $.
  Consider states $ | a_1 , a_2 \cdots a_k \rangle  $  
in  the tensor product. The action on the tensor product is 
\bea 
\sigma | a_1 , \cdots , a_k \ra = D^H_{ b_1 a_1  } ( \sigma )
  \cdots D^H_{  b_k a_k  } ( \sigma )  | b_1 , \cdots , b_k \ra
\eea 
This action of $ \sigma \in S_n$ commutes with  $S_k$ permutations
 which act 
as 
\bea 
\alpha | a_1 , \cdots , a_k \ra = | a_{\alpha (1)}  , \cdots 
, a_{ \alpha (k)}  \ra
\eea  
The symmetric subspace of $V_H^{ \otimes k }  $  is isomorphic
 to the 
space of $k$ oscillators $A_{a_1} \cdots A_{a_k }$.
We can identify $A_{a_1} \cdots A_{a_k } $
with 
\bea 
P_{sym }   | a_1 , \cdots , a_k \ra
\eea 
 where the projector is  $P_{sym} = { 1 \over k! }
 \sum_{ \alpha \in S_k } \alpha $ 
acting on  $V_H^{ \otimes k }  $.  The usual Clebsch 
decomposition gives the transformation matrix 
from the tensor product basis to a basis of irreps.
 $ | \L_1 ,i,\tau  \ra $ where $ \tau $ is a 
multiplicity index and $i$  is runs over the dimension 
$d_{\L_1 } $. 
Since $P_{sym}$ commutes with $S_n$ , its eigenvalues are constant on irreps, 
and since it is a projector they are $1$ or $0$. We can define the symmetric irreps  $ \L_1  , i$  to be the set 
 left invariant by $P_{sym}$. So we have 
\bea 
P_{sym} | \L_1  ,  i , \tau\ra =  | \L_1 ,  i  , \tau\ra  
\eea 
The Clebsch-Gordans restricted to the symmetric irreps are the symmetrised Clebsch-Gordans
\bea 
&& C_{a_1 \cdots a_k }^{ \L_1 , i, \tau  } \equiv  \la \L_1 , i , \tau  | a_1 \cdots a_k \ra  \cr 
&& =  \la    \L_1 , i, \tau   |P_{sym} |  a_1 \cdots a_k  \ra  
\eea 
If we permute the vectors in the tensor product we have 
\bea\label{symmcg}  
C_{a_{\alpha( 1 )}  \cdots a_{ \alpha (k)}  }^{ \L_1 , i, \tau  }   && =  \la   \L_1 , i, \tau  |P_{sym} \alpha  |  a_1 \cdots a_k \ra   \cr 
&& = \la     \L_1 , i, \tau  |P_{sym}   |  a_1 \cdots a_k  \ra  \cr 
&& = C_{a_1   \cdots a_k }^{ \L_1 , i, \tau   } 
\eea 
The $ \alpha $ can be absorbed in the redefinition of the summation over
permutations in $P_{sym}$.
This gives the desired symmetry of the symmetrised Clebsch
 (\ref{symmclebsch1}). 

Orthogonality (\ref{orthogsc}) follows from the restriction of the usual
 orthogonality to the symmetric subspace.  
The identity (\ref{symmclebsch2})  follows by considering 
\bea\label{symmcg2}  
 \la  \L_1 , i, \tau  | \sigma |  b_1 , \cdots , b_k \ra  
&& = \la  \L_1 , i, \tau | P_{sym}   \sigma |  b_1 , \cdots , b_k \ra \cr 
&& = \la  \L_1 , i, \tau  |    \sigma P_{sym} |  b_1 , \cdots , b_k \ra 
\eea
and inserting on the left of $ \sigma $ a complete set of tensor product states 
or on the right of  $\sigma $ a complete set of symmetric irrep states. 

\subsection{Symmetrised Clebsch multiplicities} 

We give some examples of the symmetrised Clebsch multiplicities 
for $V_H = [n-1,1]$. 
The tensor product of $ V_H \otimes V_H $ is decomposed into 
irreps as $ [ n ] + [ n-1 , 1 ] + [ n-2 , 2] + [ n-2 , 1^2] $. 
It can be checked that their dimensions add up to $ (n-1)^2 $ . 
If we drop the last we get the dimensions 
adding to $ { n ( n-1) \over 2 } $ as expected 
form the symmetric product. So we have 
\begin{equation}
\Sym (  V_H \otimes V_H ) = [ n ] + [ n-1 , 1 ] + [ n-2 , 2] 
\end{equation}

To get $ \Sym ( V_H \otimes V_H \otimes V_H ) $ we first take the
above symmetric tensor product and then a further tensor product with
$ [n-1 , 1] $.  We need the tensor product
\begin{equation}
  [ n-2 , 2]  \otimes  [ n-1 , 1 ] = [ n-1 , 1 ] + [n-2,2] + [ n-2 , 1, 1 ] + [ n-3 , 3  ]  + [n-3 , 2, 1 ] 
\end{equation}
The dimensions add correctly. Then we need to implement the symmetric
projection. One finds that the following gives the correct dimension
counting
\begin{equation}
  \Sym (  V_H  \otimes V_H \otimes V_H ) = [ n ] + 2 [ n-1 , 1 ] + [ n-2
  , 2]+ [ n-2 , 1, 1 ] + [ n-3 , 3  ]
\end{equation}
So the representation $[n-3 , 2, 1 ]$ together with one of the two
reps $[ n-2 , 1, 1 ]$, one of the two reps $[ n-2 , 2]$, and one of
the three reps $[ n-1 , 1 ]$ have been projected out from $\Sym ( V_H 
\otimes V_H )\otimes V_H$.

\section{Action of $S_n$ on $A_a^{\dagger} $ in detail}\label{sec:matrixconstruct}

The orthogonal basis of $A_a^{\dagger} $ which we used in Section 2
has the property that $S_n$ acts on it via the standard
Young-Yamanouchi orthogonal basis of the $[n-1,1]$ hook representation
(as given for example in \cite{hamermesh}).  If $s_i = (i,i+1)$ are
the 2-cycle permutations that generate $S_n$ then we have
\begin{align}
  s_iA_a^\dagger & = A_a^\dagger \quad \textrm{for } i\leq a-1
  \textrm{ and } i\geq a+2 \nn\\
  s_{i+1}A_i^\dagger & = \frac{1}{i+1} A_i^\dagger +
  \frac{\sqrt{i(i+2)}}{i+1} A_{i+1}^\dagger \nn\\
  s_{i+1}A_{i+1}^\dagger & =  \frac{\sqrt{i(i+2)}}{i+1} A_i^\dagger -
  \frac{1}{i+1} A_{i+1}^\dagger  \label{gibbon}
\end{align}

We can identify the representation of $S_n$ formed by the $A_a$ using
general arguments. It is easy to see that there is no invariant vector
under $S_n$, and that there is one invariant vector under $S_{n-1}$
(namely $A_{n-1}$). The only irreducible representations of $S_n$
which contain the invariant of $S_{n-1} $ are $[n]$ and $[n-1,1]$.
Having ruled out the symmetric irrep.  $[n]$, the $(n-1)$ dimensional
representation formed by the $A_a$ can only be the irreducible
$[n-1,1]$. More directly we can use the construction of the orthogonal
representing matrices given in \cite{hamermesh}, which uses branching
arguments.

\section{Code}\label{sec:code}

Code written to calculate the various multiplicities discussed here is
available under the \href{http://www.gnu.org/copyleft/gpl.html}{GNU
  General Public Licence} at \url{http://www.nworbmot.org/code/}.  It
is written in \href{http://www.python.org/}{python} for use with the
\href{http://www.sagemath.org/}{SAGE} open source computer algebra
system.

\end{appendix}

\end{document}